\documentclass[final,5p,times,twocolumn,preprint]{elsarticle}
\usepackage{tikz}
\usepackage{helvet}
\usepackage{color}
\usepackage{xcolor}
\usepackage{graphicx} 
\usepackage{subcaption}

\newcommand{\twCO}{$^{12}$CO}

\newcommand{\otz}{$J$=\,1--0}
\newcommand{\km}{\,{\rm km}}

\newcommand{\ps}{\,{\rm s}^{-1}}
\newcommand{\kpc}{\,{\rm kpc}}

\def\gray {$\gamma$-ray }

\def\aap{Astronomy and Astrophysics}

\def\apjl{ApJL}

\def\apjs{ApJS}

\definecolor{mydarkgreen}{RGB}{0,140,0} 

\usepackage{amssymb, amsmath}

\journal{Science Bulletin}

\begin{document}


\begin{frontmatter}
\title{An Ultrahigh-energy $\gamma$-ray Bubble Powered by a Super PeVatron}
\centerline{\author{\large LHAASO Collaboration\footnotemark[1]$^,$\footnotemark[2]}}






\begin{abstract}
We report the detection of a $\gamma$-ray {\it bubble} spanning at least 100\,$\rm deg^2$  in ultra high energy (UHE) up to a few PeV in the direction of the star-forming region Cygnus X, implying the presence Super PeVatron(s) accelerating protons to at least 10 PeV. A log-parabola form with the photon index $\Gamma (E) = (2.71 \pm 0.02) + (0.11 \pm 0.02) \times \log_{10} (E/10 \ {\rm TeV})$ is found fitting the gamma-ray energy spectrum of the bubble well.  UHE sources, ‘hot spots’ correlated with very massive molecular clouds, and a quasi-spherical amorphous $\gamma$-ray emitter with a sharp central brightening are observed in the bubble. In the core of $\sim 0.5^{\circ}$,  spatially associating with a region containing massive OB association (Cygnus OB2) and a microquasar (Cygnus X-3), as well as previously reported multi-TeV sources, an enhanced concentration of UHE $\gamma$-rays are observed with 2 photons at energies above 1 PeV. The general feature of the bubble, the morphology and the energy spectrum, are reasonably reproduced by the assumption of a particle accelerator in the core,   continuously injecting protons into the ambient medium.

\end{abstract}




\end{frontmatter}




\footnotetext[1]{Corresponding authors:  Zhen Cao (caozh@ihep.ac.cn), Cong Li (licong@ihep.ac.cn),  C.D. Gao (gaocd@ihep.ac.cn), R.Y. Liu (ryliu@nju.edu.cn), R.Z. Yang (yangrz@ustc.edu.cn)  }
\footnotetext[2]{The LHAASO Collaboration authors and affiliations are listed in the supplementary materials}

\section{Introduction}
\label{}
Cygnus-X  is one of the most intensive and nearby (at a distance $\approx 1.4$~kpc) star-forming regions in the Milky Way. The $7^{\circ}\times7^{\circ}$ area harbors several Wolf-Rayet stars and hundreds of O-type stars grouped in powerful OB associations. It also contains vast HI and molecular gas complexes with masses exceeding  $10^{6} M_\odot$.
The presence of potential particle accelerators, namely massive stars and supernova remnants, and targets for $\gamma$-ray production (dense gas regions) make several parts of this region effective $\gamma$-ray emitters. Fermi-LAT has detected high energy $\gamma$-rays from the direction of the massive star association Cygnus OB2. The extended $\sim$$2^\circ$ source (dubbed `Cygnus Cocoon'\cite{Ackermann:2011}) later has been reported  at TeV\cite{Bartoli:2014,HAWC-Cocoon} and 
PeV photo was also detected from this direction \cite{LHAASO-12-Pevatrons} energies. 
The latter has attracted particular attention due to its potential connection with the origin of the highest energy galactic CRs around and beyond the so-called `knee'.  Here, we present the results of new observations of the Cygnus region with the Large High Altitude Air Shower Observatory (LHAASO) and discuss their astrophysical implications.


LHAASO is a mega-scale dual-task facility designed to study cosmic rays and $\gamma$-rays. It is installed at 4410~m above sea level in Sichuan Province, China\cite{2010ChPhC..34..249C}. In the $\gamma$-ray detection mode, the LHAASO detectors cover over three energy decades. The 78,000 m$^2$ Water Cherenkov Detector Array (WCDA) operates below 10~TeV down to 100 GeV, while the Kilometer Square Array (KM2A), consisting of 5216 surface scintillator detectors and 1188 underground muon detectors, is optimized for the band of highest energy $\gamma$-rays 
from 10~TeV to a few PeV \cite{he18}. The LHAASO detectors cover a significant part of the sky spanning in declination between   -21$^\circ$ and  +79$^\circ$.  While the surface detectors provide adequate $\gamma$-ray photon statistics,  thanks to the effective rejection of the hadronic (CR-induced) showers, the minimum detectable energy flux approaches a level unprecedented in $\gamma$-ray astronomy,  $\sim 10^{-14} \, \rm erg/cm^2 s$. Depending on the photon energy, the angular resolution varies from $0.5^\circ$ to $0.2^\circ$. Above 100 TeV, the energy resolution is better than 20 per cent.  
The LHAASO array was completed in June 2021, although the data taken since 2019 with partly installed detectors revealed a dozen CR PeVatrons in the Galactic Disk \cite{LHAASO-12-Pevatrons}. The detection of a 1.4 PeV photon from the Cygnus region was 
a big surprise, making this source one of the highest-priority targets for observations. LHAASO is ideally located for tracking the Cygnus region, providing more than 2,000 hours of exposure time per year. 

\section{The LHAASO Observation}

Our new observations firstly confirm the existence of a very extended emission at Cygnus region in ultra-high energy band.
The discovery beyond our initial publication\cite{LHAASO-12-Pevatrons} includes a few striking findings, namely  (i) the detection of an enormous $\gamma$-ray bubble that extends to more than 6$^\circ$ from the center, (ii) the detection of hot spots associated with massive molecular clouds, (iii)  the observation of the SED of the bubble up to 2 PeV, and (iv) finding a new source nearly at the center of the bubble, spatially coincident with both Cygnus OB2 and Cyg X-3. 


The new results are based on substantially enhanced (by a factor of five) statistics of photons detected by KM2A and WCDA since the beginning of full operation in 2021. Overall, about 3200 photon-like events with energies over 100 TeV have been registered  from LHAASO J2032+4102 within the radius 10$^{\circ}$. 
Moreover, thanks to WCDA, the spectral measurements have been extended down to 2~TeV. The new data set allows spectral and morphological measurements with comparable angular and energy resolutions over three energy decades.   


\begin{figure*}[htbp]
\centering
\includegraphics[width=0.9\linewidth]{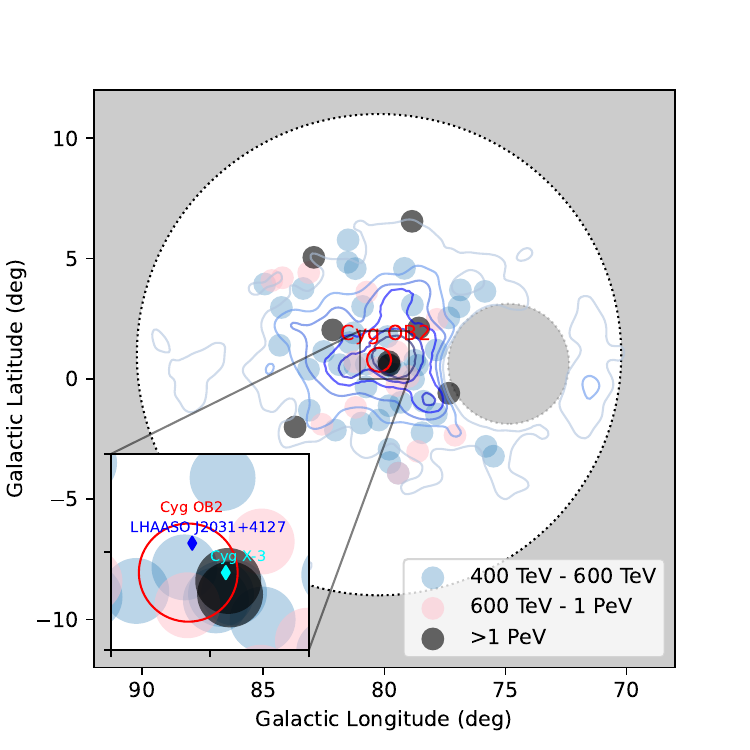}
\caption{
Photon distribution in the Cygnus-X region. The size of the circle labels the point spread 
 function of LHAASO-KM2A in the energy range above $100~\rm TeV$. The significance map of $\gamma$-rays from $2 ~\rm TeV$ to $20~\rm TeV$ of the bubble is shown by grey contours starting from 3$\sigma$ with a step of 3$\sigma$. This structure is about 20$^\circ$ in longitude and latitude. The blue diamond located at the centre of the $\gamma$-ray image marks LHAASO J2031 + 4127,
 which coincides with the unidentified source  TeV~2031+4130 \cite{hegra_cygnus}. There are 66 photon-like events within a radius of 6 degree with an estimated background of 9.5. Eight events with energy above $1 ~\rm PeV$ are marked with black circles, 12 events with energy between $600~\rm  TeV$ and $1 ~\rm PeV$ are shown as pink,  and the other 46 events with energy between $400~\rm  TeV$ and $600~\rm  TeV$  are shown with blue circles. The photons above 400~TeV extend beyond  6$^\circ$, but with a higher CR background contamination, so we do not show them individually on the map. In particular,  as shown in the zoom-in figure, seven  of these high energy photons, 2 of them with energy above 1\,PeV, are located in the region of radius of 0.5$^\circ$ relative to the centre  (red circle), which is roughly the size of the massive star association Cygnus OB2. Possible contamination of the CR background is only 0.07 events.  This region contains at least three interesting objects -  Cygnus OB2 (red circle), Cyg X-3 (cyan diamond) and the powerful pulsar PSR J2031+4127 (blue diamond). The larger circle in black dotted line represents the ROI used in this study, while the shaded circle within black dotted line marks the  masked region near the unidentified  source LHAASO J2018+3651.  This source is bright and reveals a large spatial extension, thus a circular region with a radius of $2.5^{\circ}$ is  masked in the analysis. 
}
\label{Fig:400TeV-bubble}
\end{figure*}

Figure \ref{Fig:400TeV-bubble} shows the contour map of  2 to 20 TeV $gamma$-rays obtained towards  Cygnus X, 
and 66 individual photon-like events of energies exceeding 400~TeV. The estimated cosmic ray background is about 9.5, and the contamination will be smaller with the increase of energy benefiting from the improvement of rejection power. There are 8 events with energy above 1 PeV, whileas the background is only 0.75. Considering the small background-to-signal ratio, 
we get valuable information from the individual events. They are distributed inhomogeneously. In general, the density is higher at a closer distance to the core. Especially within 0.5$^{\circ}$ radius  
around Cygnus  OB2, 7 events above 400~TeV have been detected, including two photons of energies exceeding 1 PeV. 
This implies the operation of central CR accelerator(s) that injects relativistic protons and nuclei into the circumstellar medium (see Supplementary Section).  


To study the hot spots and the much broader diffuse $\gamma$-ray structure (`bubble'), the contribution from all individual sources that show up in the region of interest (ROI) should be removed from the analysis. To separate the signal from individual sources and extended emission, a 3-dimensional(`3-D') fitting procedure has been developed, which is widely used by experiments  in the $gamma$-ray band to deal with complex astronomical environments. The spatial and spectral parameters are fitted simultaneously by maximizing the likelihood value.
The very extended emission is modeled by a combination of a Gaussian distribution and a template based on the angular pattern of the HI and ${\rm H}_2$ gas distributions. 

\subsection {Individual sources}
In this relatively compact central region with a radius of $0.5^\circ$ (hereafter, the core) 
are located Cygnus OB2, a  massive OB association, and the powerful X-ray binary Cyg X-3. There are three TeV sources, including the the hard-spectrum TeV source TeV J2032+4130 \cite{hegra_cygnus} (=LHAASO J2031+4127), binary system PSR J2032+4127/MT91
213\cite{Abeysekara_2018} and the ultra-high energy source LHAASO J2032+4102\cite{LHAASO-12-Pevatrons}, are detected in this region.
The resolved sources potentially correlated with previous TeV sources are discussed in supplementary material. A new source (LHAAO J2031+4057), detected by WCDA below 10 TeV,   is also found in the core region, in  spatial association with both Cygnus OB2 and Cyg X-3.  
The similarity of spectral shape with the extended emission suggests that this source is presumably a part of the bubble. The increase of TS by adding a new source at the core region is 18 for KM2A, which shows a hint of excess but it is not significant now.

Immediately outside the {core} is a TeV $\gamma$-ray source, which is associated with the middle-aged supernova remnant $\gamma$-Cygni \cite{magic_gcygni}. The source has been detected by LHAASO  up to 100~TeV. The detailed study of this interesting object, which overlaps with the bubble, will be published elsewhere. In this paper, it is modeled using a 2D Gaussian template. The $\sigma$-parameter of the Gaussian template is 0.23$^\circ$.



\subsection{The Cygnus Bubble}

\begin{figure*}[htbp]
  \centering
  \begin{tikzpicture}
    \node at (0-1.8+0.7,0+2.2) {\includegraphics[width=0.3\textwidth]{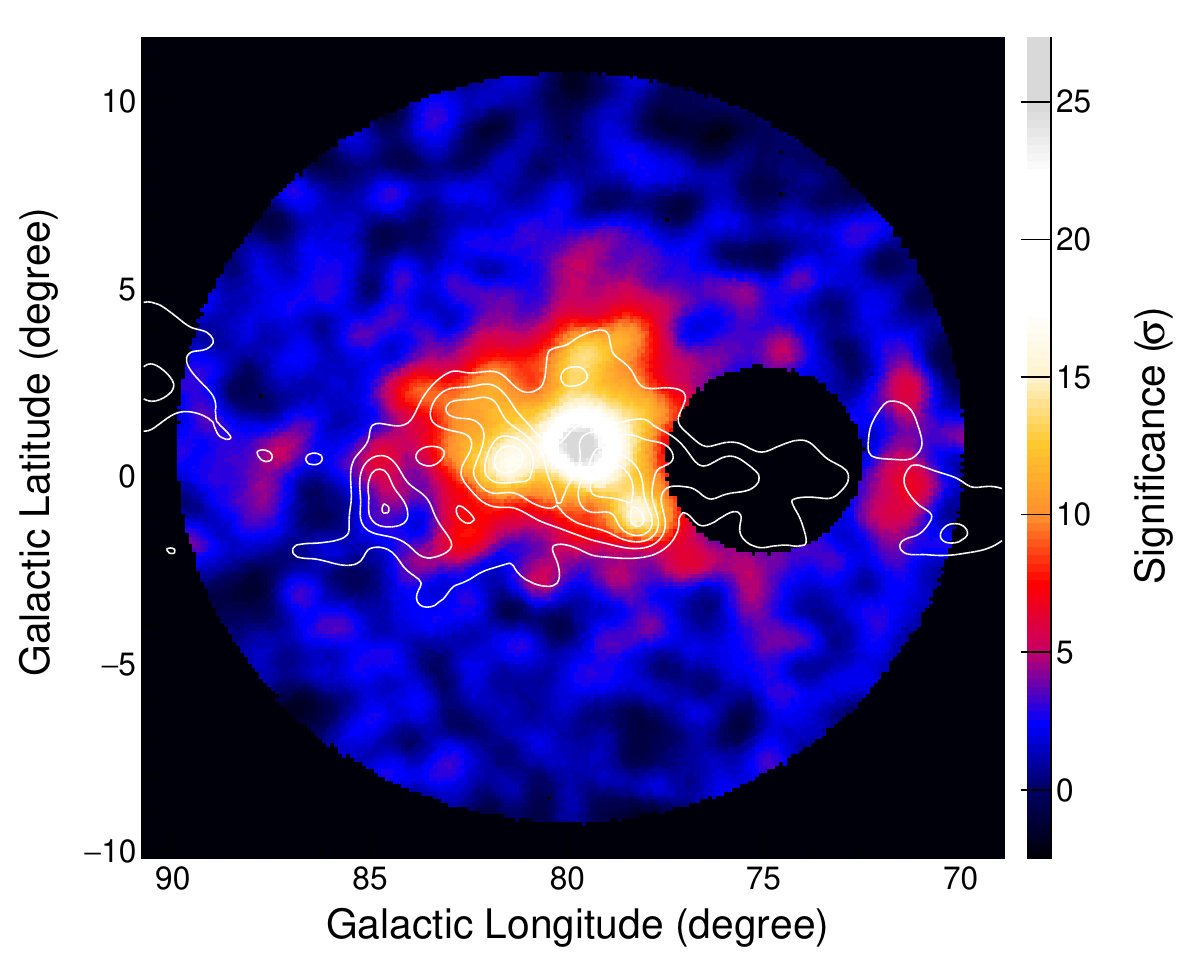}};
    \node at (4,0+2.2) {\includegraphics[width=0.3\textwidth]{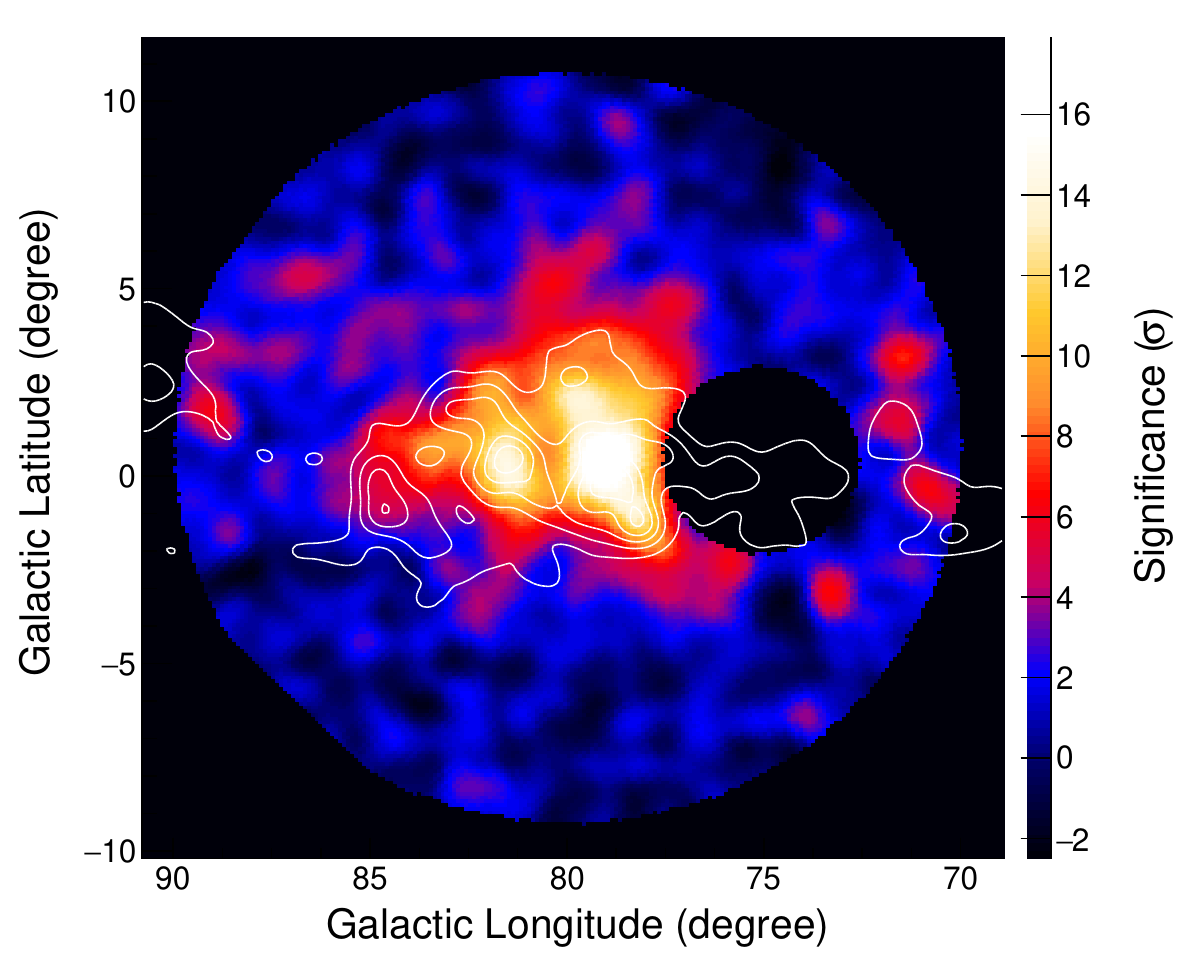}};
    \node at (8+1.8-0.7,0+2.2) {\includegraphics[width=0.3\textwidth]{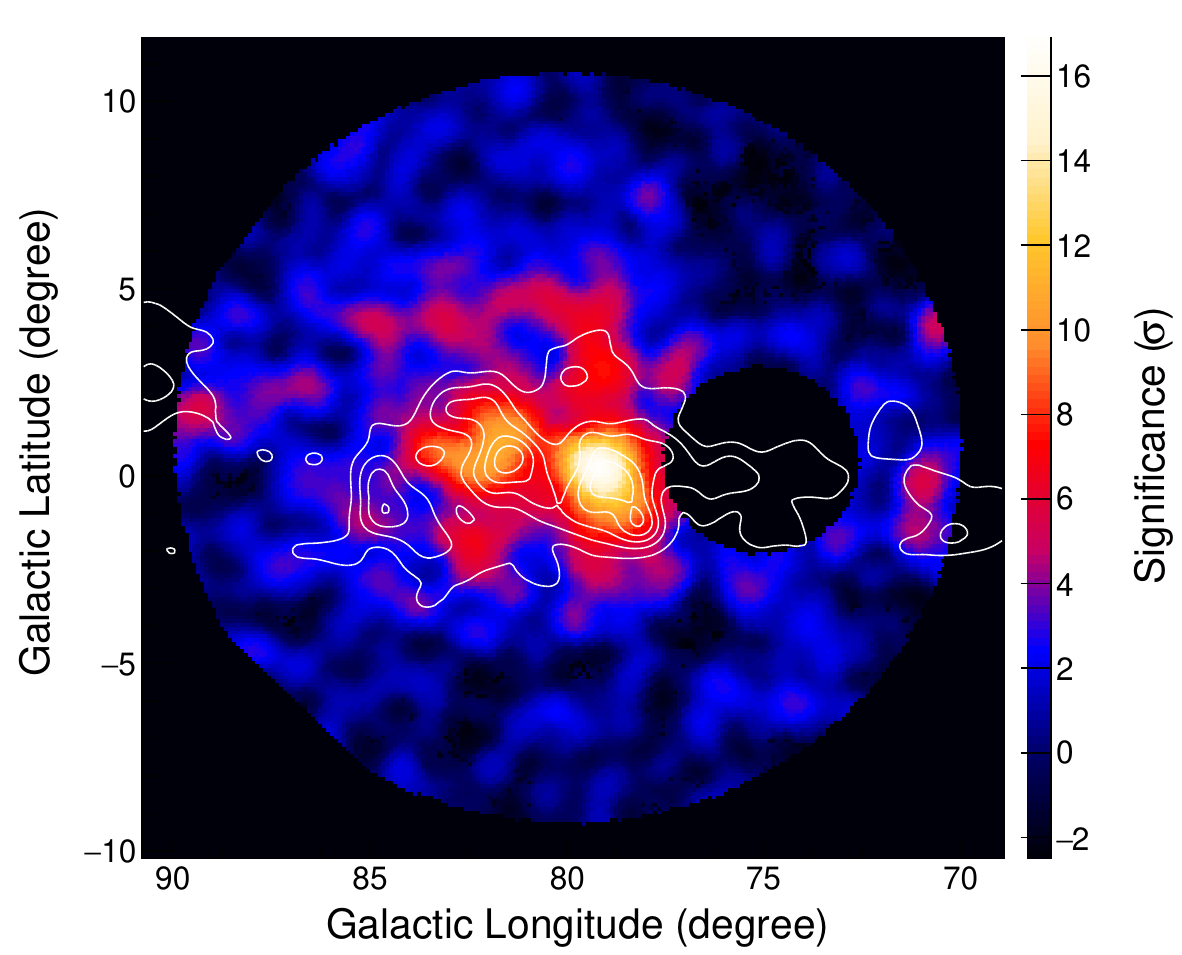}};
    \node at (0-1.8+0.7,-3.5+0.5+0.5) {\includegraphics[width=0.3\textwidth]{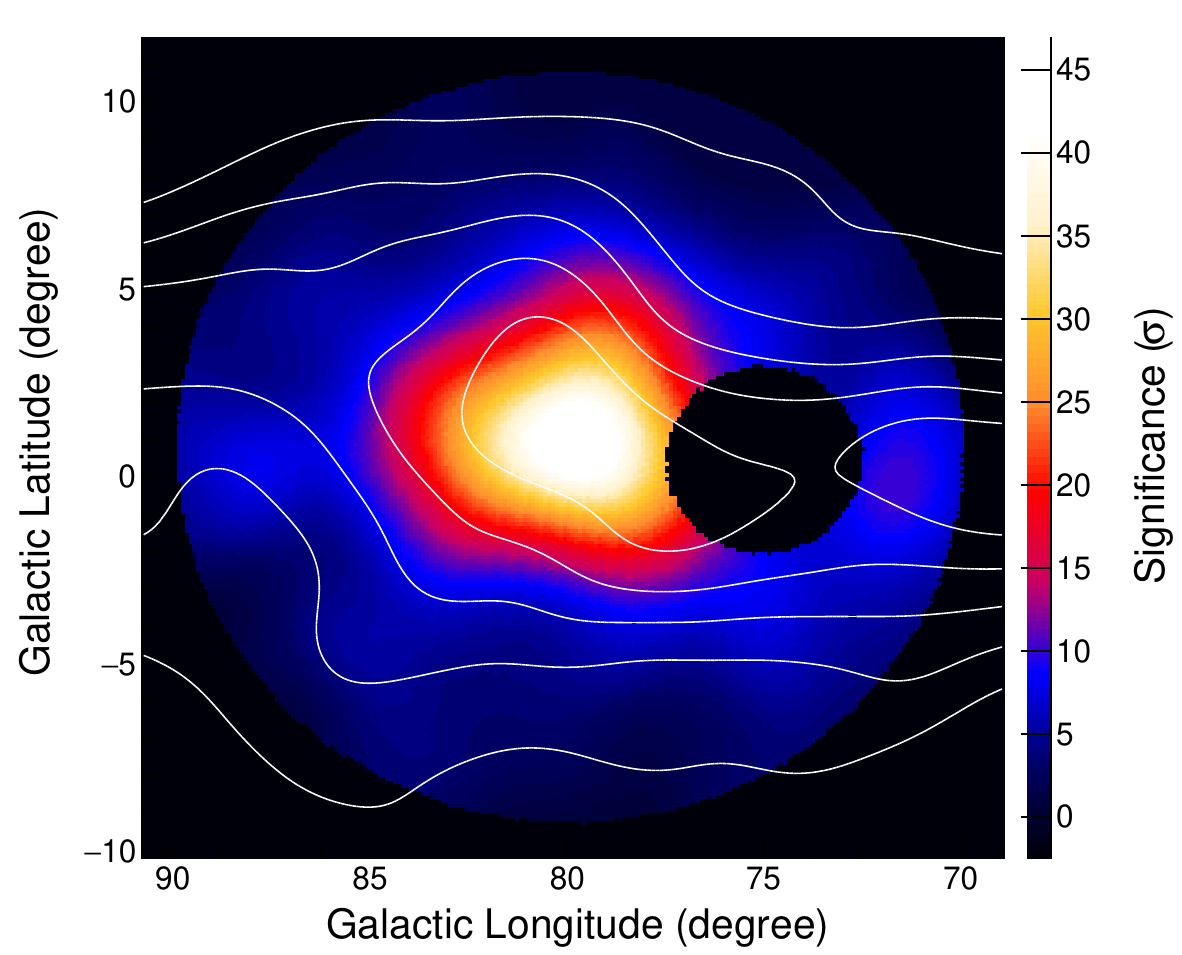}};
    \node at (4,-3.5+0.5+0.5) {\includegraphics[width=0.3\textwidth]{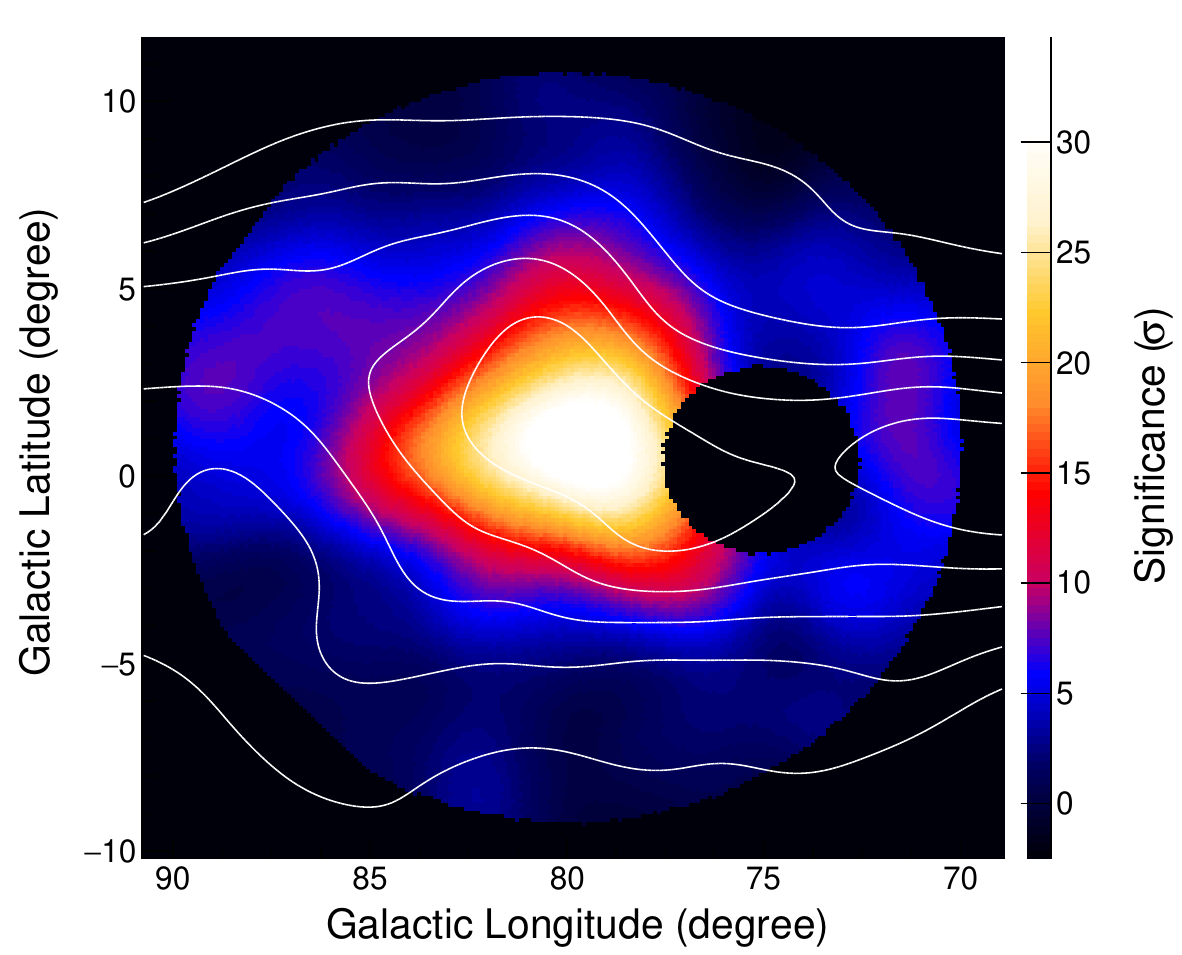}};
    \node at (8+1.8-0.7,-3.5+0.5+0.5) {\includegraphics[width=0.3\textwidth]{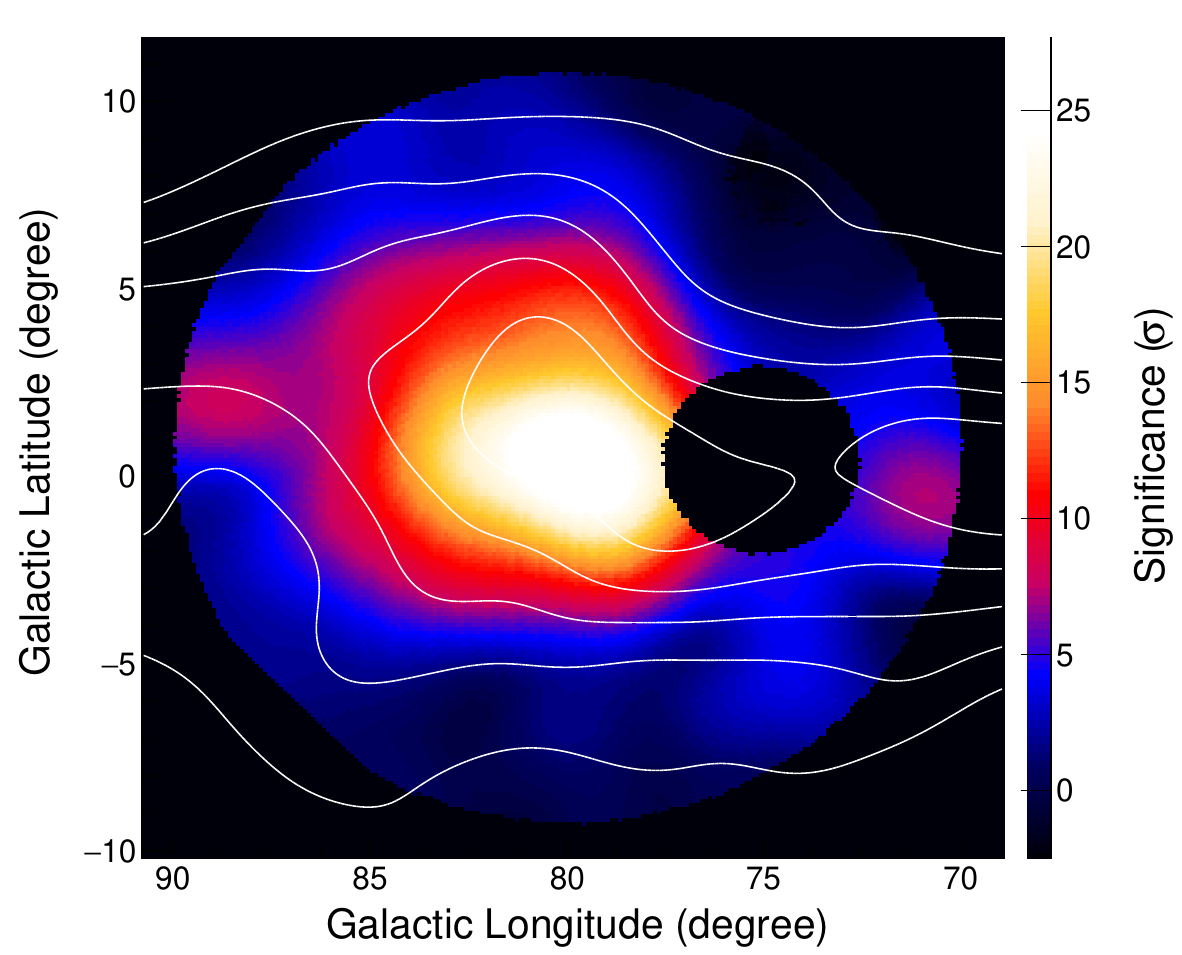}};
    \node at (-0.05-0.074-1.8-0.05+0.55+0.01+0.15,-8.2+1.1) {\includegraphics[width=0.27782609\textwidth]{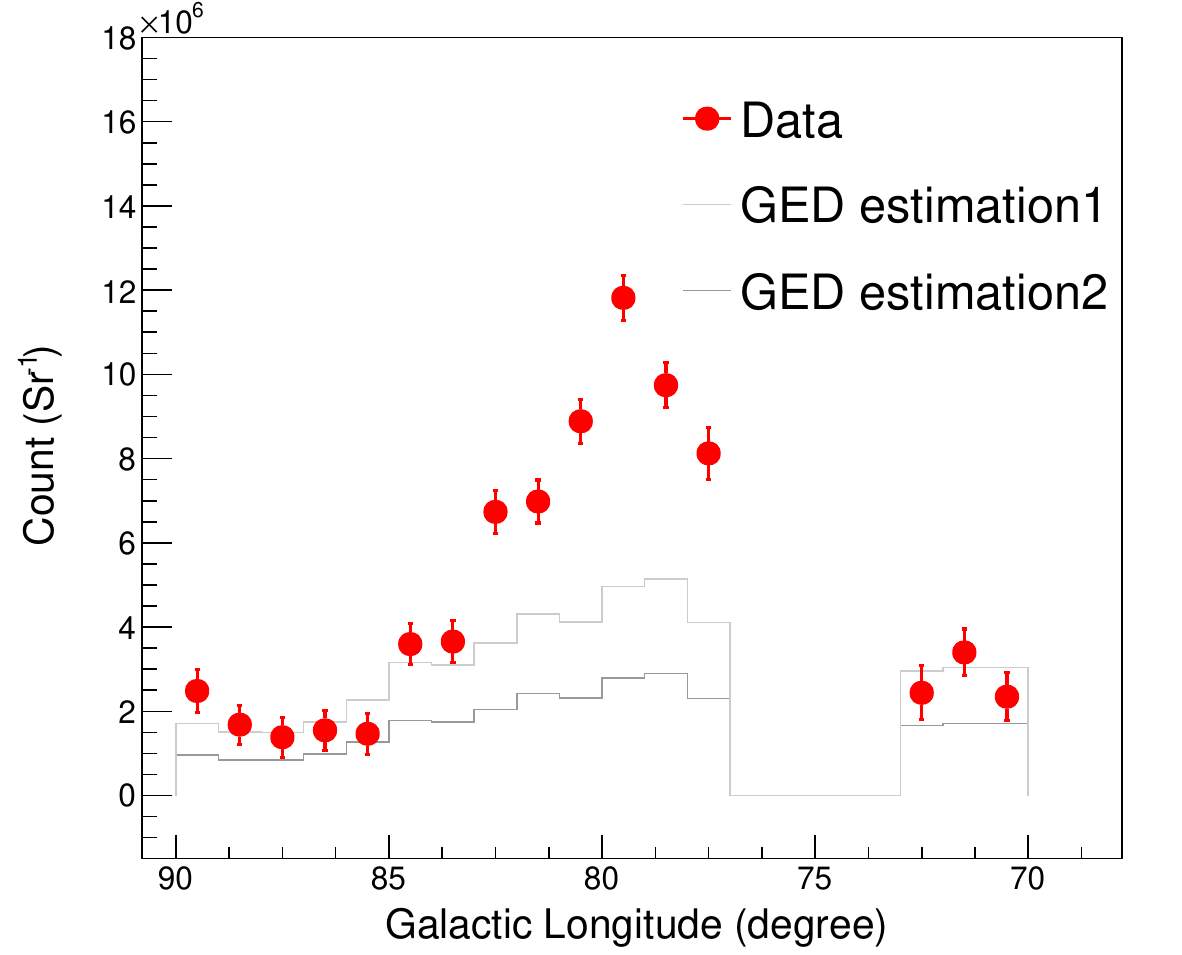}};
    \node at (3.95-0.074-0.05+0.01,-7.+0.05-2.2+0.95+1.1) {\includegraphics[width=0.27782609\textwidth]{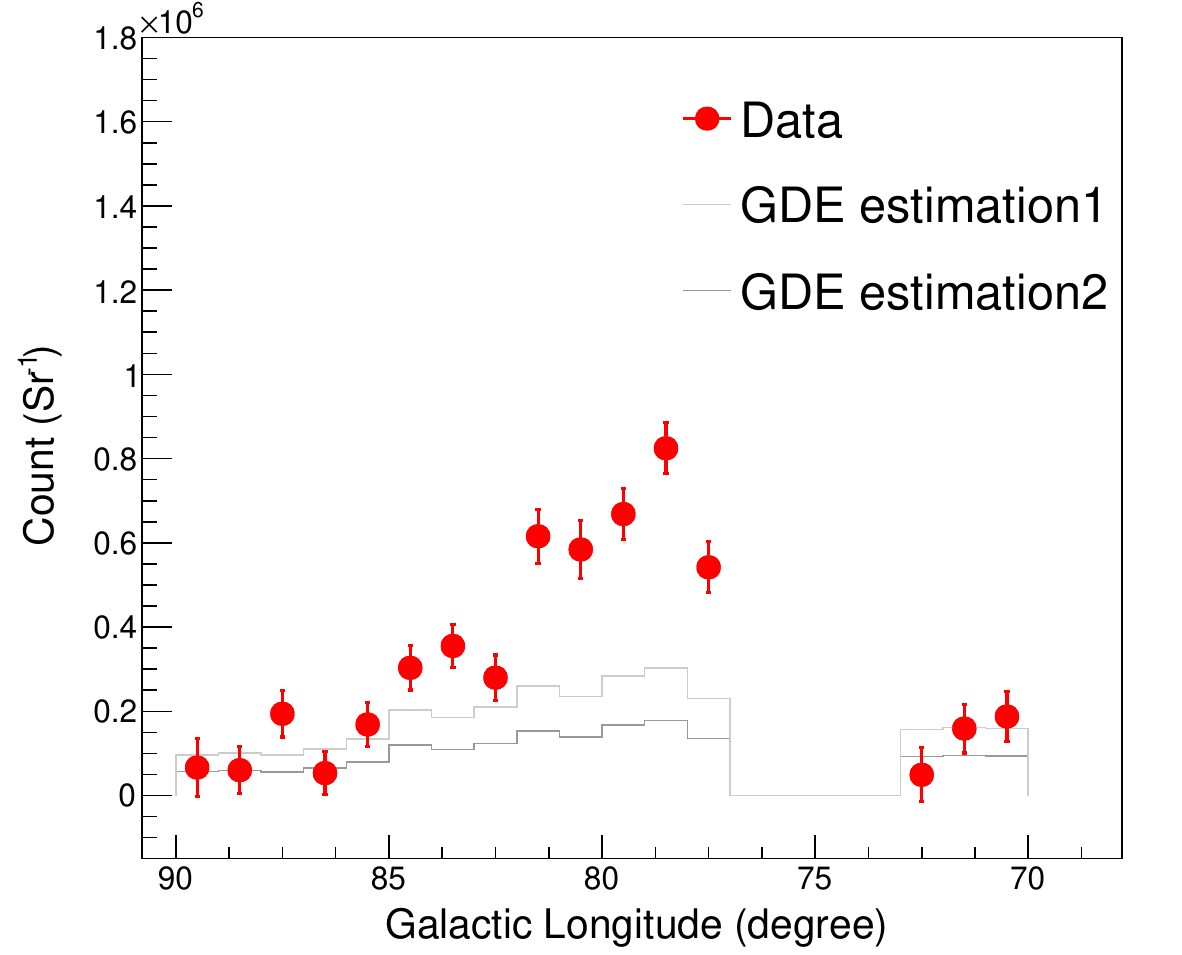}};
    \node at (7.95-0.074+1.8-0.05-0.55+0.01-0.15,-7.+0.05-2.2+0.95+1.1) {\includegraphics[width=0.27782609\textwidth]{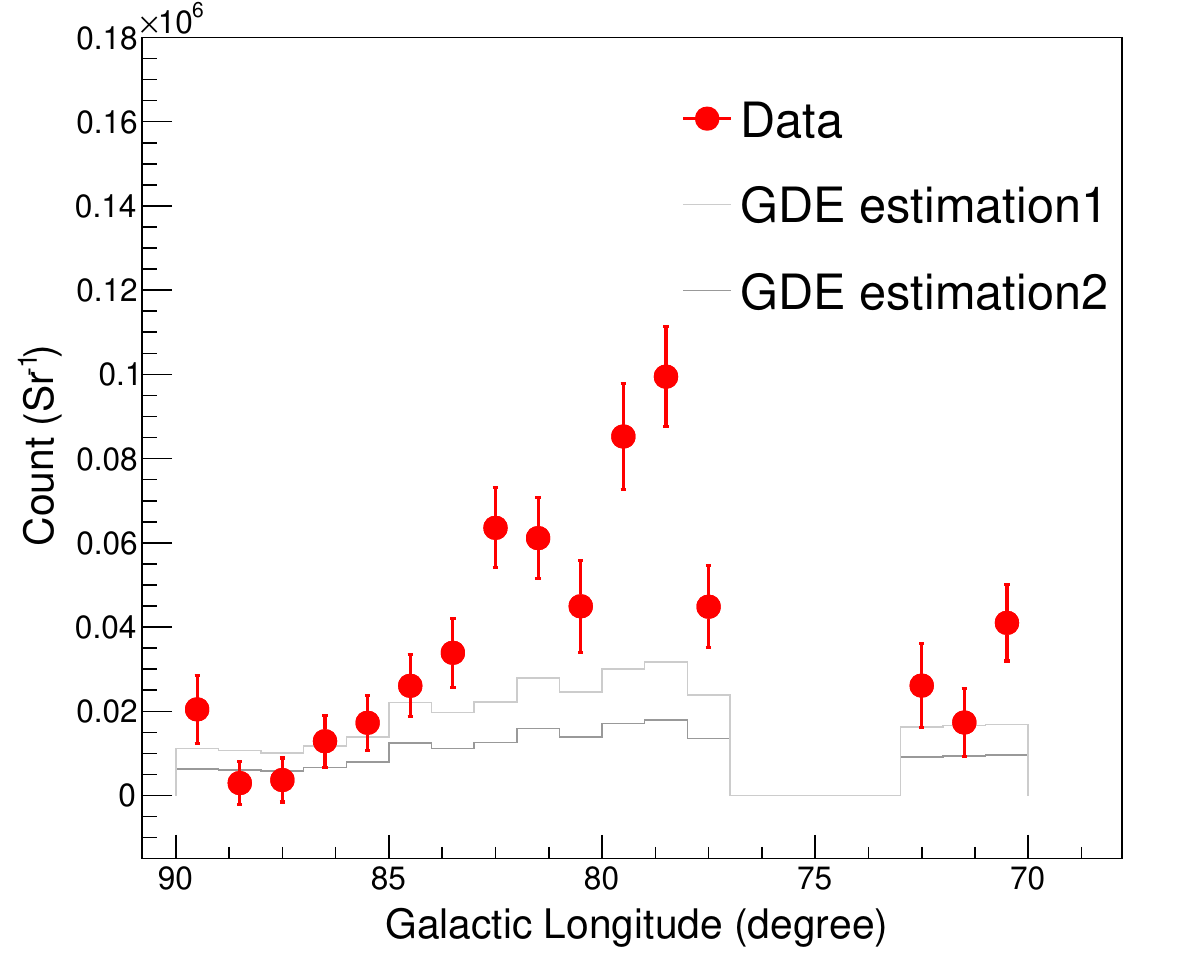}};
    
    \node[above, font=\fontfamily{phv}\fontsize{13pt}{28pt}\selectfont\color{mydarkgreen}] at (-0.1+0.4+0.15,1.+2.6+0.1-0.2) {A};
    \node[above, font=\fontfamily{phv}\fontsize{13pt}{28pt}\selectfont\color{mydarkgreen}] at (5.55,1.+2.5) {B};
    \node[above, font=\fontfamily{phv}\fontsize{13pt}{28pt}\selectfont\color{mydarkgreen}] at (10.8-0.15,1.+2.5) {C};
    \node[above, font=\fontfamily{phv}\fontsize{13pt}{28pt}\selectfont\color{mydarkgreen}] at (0.3+0.15,-1.2) {D};
    \node[above, font=\fontfamily{phv}\fontsize{13pt}{28pt}\selectfont\color{mydarkgreen}] at (5.55,-1.2) {E};
    \node[above, font=\fontfamily{phv}\fontsize{13pt}{28pt}\selectfont\color{mydarkgreen}] at (10.8-0.15,-1.2) {F};
    \node[above, font=\fontfamily{phv}\fontsize{13pt}{28pt}\selectfont\color{mydarkgreen}] at (0.3+0.15,-5.8) {G};
    \node[above, font=\fontfamily{phv}\fontsize{13pt}{28pt}\selectfont\color{mydarkgreen}] at (5.55,-5.8) {H};
    \node[above, font=\fontfamily{phv}\fontsize{13pt}{28pt}\selectfont\color{mydarkgreen}] at (10.8-0.15,-5.8) {I};
  \end{tikzpicture}
  \caption{The Cygnus Bubble in 3 decades of photon energy. Two-dimensional significance maps of the Cygnus Bubble in the region of interest, which are smoothed with a Gaussian kernel  of $\sigma$=0.3$^{\circ}$ (upper row) and $\sigma$=1$^\circ$ (middle row). 
  All individual sources including the SNR $\gamma$-Cygni are removed.
  One-dimensional angular distributions of photons in the bubble are displayed in the lower row. From left to right, the columns of panels are for photon energies in the ranges of 2-20 TeV, 25-100 TeV and above 100 TeV, respectively. The structures of the bubble at different energies are visible in the upper panels (A, B and C). Hot spots are revealed at energies above 25 TeV associated with the local molecular cloud distribution, which is indicated by the contours. The broad structure of the bubble in the middle row of panels (D, E and F) has a good association with the local HI gas distribution, which is indicated by the contours. This structure covers a very wide region i.e, 10$^\circ$ from the core. The distribution of $\gamma$ emission in Galactic longitude with a latitude range from $-2^{\circ}$-$2^{\circ}$. The two gray lines are the estimated diffuse emission from two different regions.The GDE estimation 1 and 2 are derived from inner and outer galaxy region separately.}
  \label{Fig:hot-spots-bubbles}
\end{figure*}

After the removal of all (identified and unidentified)  $\gamma$-ray sources and `hot spots', a giant  $\gamma$-ray structure (hereafter, the Cygnus Bubble) reveals both in the WCDA and KM2A data. The residual structure, detected from $\sim$2~TeV to $\geq 1$~PeV and spreading to $\sim 10^\circ$, formally is fitted with a Gaussian template for the inner part together with an extended tail approximately proportional to the HI angular distribution pattern. The values of the $\sigma$-parameter of the Gaussian distribution derived from the WCDA and KM2A data  are similar:  $2.28^\circ\pm0.14^\circ$ and $2.17^\circ\pm0.10^\circ$, with centers at $(l,b)=(79.61^\circ\pm0.23^\circ$, $1.65^\circ\pm0.24^\circ$) and ($79.62^\circ\pm0.18^\circ$, $1.16^\circ\pm0.18^\circ$), respectively. 
The significance maps in different energy intervals show a strong brightening of the bubble toward the core
(see Figure \ref{Fig:hot-spots-bubbles}). 

The distribution of the excess counts and the estimated Galactic diffuse $\gamma$-ray emission(GDE) as functions of the galactic longitude are shown in Figure~\ref{Fig:hot-spots-bubbles} (the bottom row of panels). 
It is clear that the $\gamma$-ray brightness distribution is much sharper than the distribution of GDE.
This excludes a significant contribution of GDE 
to the flux coming from the bubble itself. Indeed, the GDE is produced in interactions of the cosmic ray `sea' with the interstellar gas throughout the Galaxy, which is proportional to the total gas column density in the given direction,
$N_{\rm HI+H_2}(\theta)$.
The measured $\gamma$-ray fluxes substantially exceed, by a factor of 2-3, any realistic estimate of the GDE flux (see below). 
The bubble-to-GDE ratio clearly increases towards the center of the bubble.

In principle, the angular distribution of the bubble emission could be a result of the superposition of a very large number of unresolved discrete $\ gamma$-ray sources, but such a concentration of UHE $\gamma$-ray emitters in a relatively small volume hardly could be the case. More realistically, it seems to be linked to the local cosmic rays, i.e., relativistic protons and nuclei accelerated inside the bubble and distributed with enhanced density in the inner parts of the bubble.  

If  $\gamma$-rays are produced by CRs interacting with the ambient gas, the $\gamma$-ray morphology should correlate with the gas distribution. The $\gamma$-ray flux is proportional to the product of the CR and gas densities, 
$w_{\rm CR} \times N_{\rm HI+H_2}$, where $w_{\rm CR}$ is the energy density of CRs, thus an ideal correlation is achieved only in the case of a homogeneous distribution of CRs. Otherwise, the level of correlation is always less than 100 \%. High-quality $\gamma$-ray morphology, combined with good knowledge about the gas distribution, allows direct extraction of the spatial distribution of CRs, which in turn contains information about the location of the accelerator and the injection regime of particles into the circumstellar medium.   
Here we use the gas distributions taken from the HI4PI 21-cm line survey\cite{HI4PI} and the CfA galactic CO survey \cite{Dame01} archival data as discussed in the supplementary section. As we can see from Figure \ref{Fig:hot-spots-bubbles}, the $\gamma$-ray brightness partly correlates with the diffuse HI gas distribution. 
On top of the amorphous $\gamma$-ray component associated with the HI gas, a second component, clearly associated with dense, massive molecular clouds, is seen in the first row of Figure \ref{Fig:hot-spots-bubbles}.

To measure the flux of $\gamma$-rays produced in the bubble,  the contamination introduced by GDE should be subtracted. The latter could be significant, especially at the periphery of the bubble.  Recently, LHAASO-KM2A has measured GDE from two different regions of the Galactic plane \cite{lhaaso_diffuse}. The reported  GDE flux in the inner Galaxy is at least 2-3 times higher than the predicted flux assuming a homogeneous `Cosmic Ray Sea', while the GDE in the outer Galaxy is almost consistent with predictions. 
Here, the GDE in the IOR is estimated by re-scaling according to  the gas column densities in corresponding directions. The estimated GDE distributions \cite{lhaaso_diffuse} from two directions are shown in Figure \ref{Fig:hot-spots-bubbles}.  The origin of the enhanced GDE fluxes towards the inner Galaxy is under debate. As demonstrated below,  the size of Cygnus Bubble may extend to several hundreds of pc from the center, which is beyond the radius of the mask region used in the GDE measurements\cite{lhaaso_diffuse}.
Thus, it is possible that the GDE measured in the inner Galaxy contains significant components consisting of the superposition of 
large-scale structures (like the Cygnus Bubble) near the CR accelerators. 
Nevertheless, even if we apply the measured GDE flux from the inner Galaxy to the Cygnus region, there is a clear excess within a radius of about $6^{\circ}$ corresponding to the physical size of about $150~\rm pc$,  as shown in the bottom row of Figure \ref{Fig:hot-spots-bubbles}. Such conservative treatment of the GDE flux makes the estimation of the size of the bubble very robust. 



\begin{figure}
\centering
\includegraphics[width=0.5\textwidth,height=0.5\textwidth]{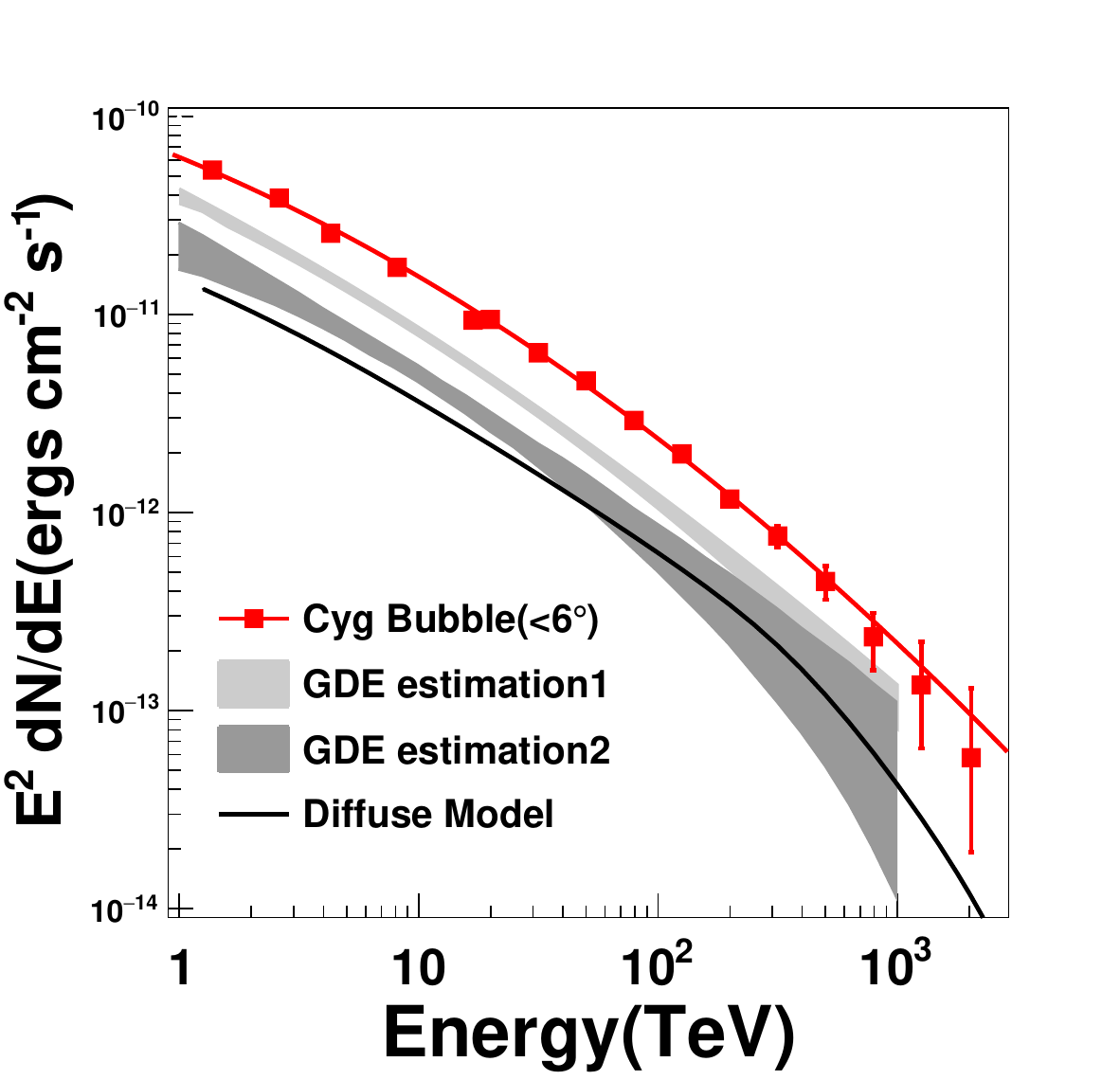}
\caption{The $\gamma$-ray flux detected toward Cygnus X integrated over the region of  radius of $6^{\circ}$.
The spectral points (red symbols) is fitted well by a log-parabola functional form with the 
photon index  $\Gamma=(2.71\pm 0.02)+(0.11 \pm 0.02)\times \log_{10}(E/10\rm TeV)$. 
The black curve corresponds the predicted flux of diffuse galactic emission calculated under the assumption that 
CRs are homogeneously distributed over the Glactic Plane.   The fluxes of the Galactic Diffuse Emission (GDE) based on the measurements from specific parts of the inner and outer parts of Galaxy  (GDE 1 and 2, respectively) 
but re-scaled for the column density toward the Cygnus Bubble are plotted as well. 
}
\label{Fig:cen-bubble-SED}
\end{figure}


The Spectral Energy Distribution (SED) of $gamma$-rays  integrated over the region with an angular radius $6^\circ$ is shown in Figure \ref{Fig:cen-bubble-SED}. {\color{red}} 

The noticeably curved spectrum is well described by a log-parabola form.   The photon index (the local power-law slope), $\Gamma (E) = (2.71 \pm 0.02) + (0.11 \pm 0.02) \times \rm log_{10} (E/10 \ TeV)$,  gradually increases with energy at least up to 2 PeV without indicating a sharp cutoff. The good agreement between the WCDA and KM2A data around 20 TeV allows a critical cross-check between different methods of spectral measurements. The measured  (red) points characterize the fluxes of the bubble, but they are not free of pollution caused by   GDE. The latter contains large uncertainties, which are transferred to the genuine flux of the bubble. 

In Figure \ref{Fig:cen-bubble-SED}, we show the calculated diffuse $gamma$-ray flux (black curve) under the assumption that they are produced by interactions of the homogeneous CR sea to be representative of the locally measured CR flux. These calculations have nonnegligible uncertainty related to the direct measurements of CR fluxes, especially around the CR knee at $\sim$1 PeV \cite{Lipari18}. However,  the latter cannot explain the large difference (by a factor of 5)  between the flux measured toward the bubble and the estimated diffuse flux. 

On the other hand, the real GDE could be significantly higher because of the strong gradient of the CR distribution in the Galactic Plane or due to the superposition of unresolved $gamma$-ray sources,  e.g., giant $gamma$-ray halos surrounding the electron and proton  PeVatrons. In Figure \ref{Fig:cen-bubble-SED},  we show two versions of GDE estimations based on the LHAASO measurements of two different segments of the inner and outer Galaxy. The measured flux toward the bubble exceeds, at least twice,  either one the estimated from the GDE measurements. This implies that the bubble contributes most of the flux detected towards 
Cygnus X. This conclusion is strongly supported by the longitudinal distribution of the $gamma$-ray flux 
(see Figure \ref{Fig:hot-spots-bubbles}). 

The energy spectrum of the `hot spots'  associated with the molecular gas is fitted as a power law with the index $\Gamma=2.86 \pm 0.04$. It is somewhat harder than the spectrum of the entire bubble, although the large statistical errors do not allow a strong statement regarding the spectrum of the hot spots.  
 The natural link between the core region and the extended bubble is that the CRs are produced inside the core by a super PeVatron, and continuously injected into  the circumstellar medium. Then, they diffuse out to fill and illuminate the entire bubble. The good correlation of the hot spots with the dense molecular gas further support this scenario.  

\section{Interpretation}

Despite the complex structure of the bubble, both the $\gamma$-ray morphology and the SED of the bubble over six decades, from 1 GeV to 1 PeV, can be interpreted within a  rather natural scenario, namely the injection of relativistic protons and nuclei from PeVatron(s) located in the core (see the Supplementary Section). Gamma-rays are produced in interactions of these particles propagating diffusively through the circumstellar gas to at least 150~pc. 
We assume that energetic protons are injected at a constant rate over the age of the system $t_{\rm age}$ from a point-like source located at the bubble center.
If Cygnus OB2 is the source of these energetic protons, we have $t_{\rm age}=2-3\,$Myr. 
%

\begin{figure*}
    \centering
    \includegraphics[width=0.8\textwidth]{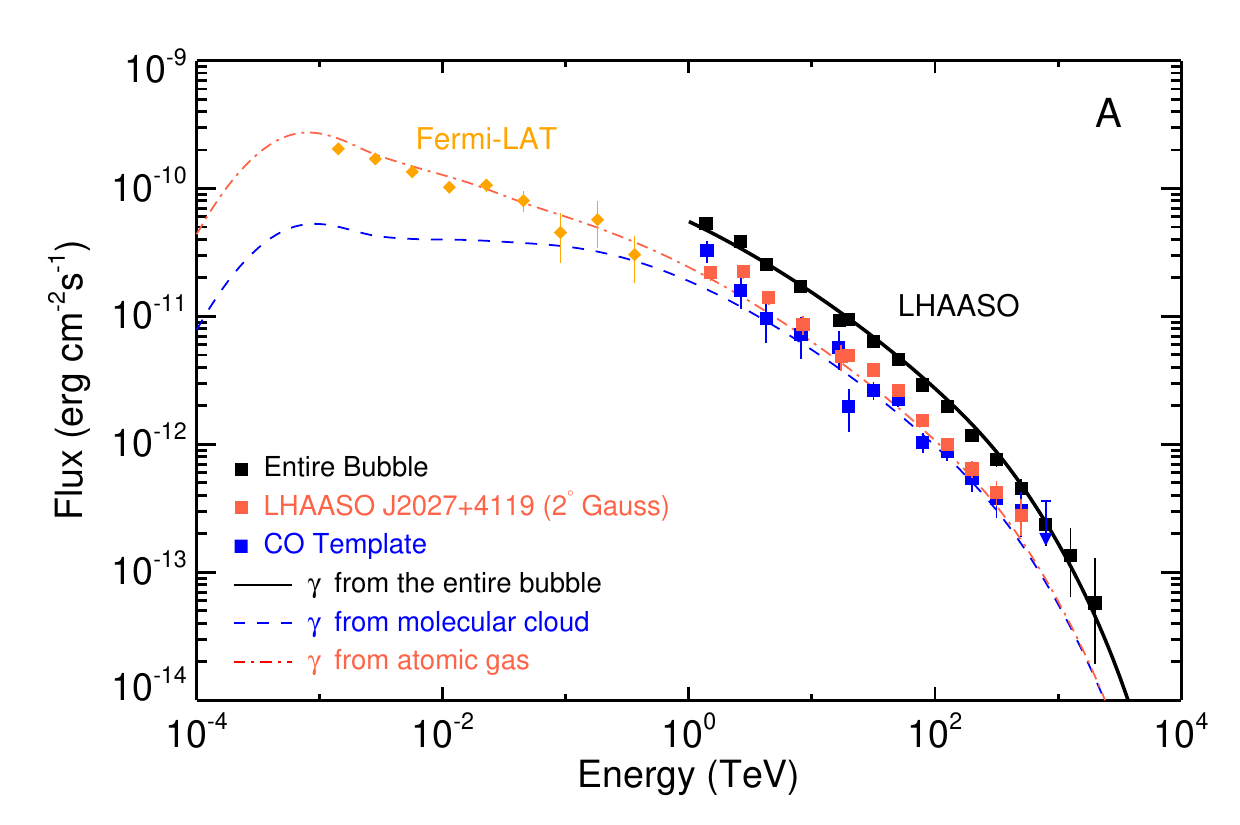}\\
   ~~~ \includegraphics[width=0.75\textwidth]{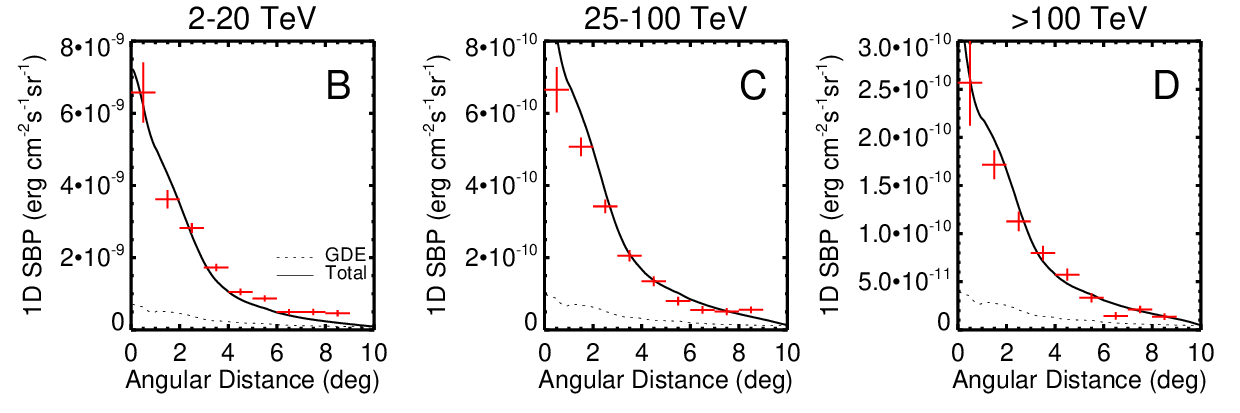}
    \caption{Modeling of the Cygnus Bubble that simultaneously fits the SEDs and 1-dimensional intensity profiles of $\gamma$-rays. In panel A, the measured fluxes from
    the entire bubble (black squares), from the $2^\circ$ Gaussian component or LHAASO~J2027+4119 (red squares), and from the CO template (blue squares). Orange diamonds present the flux of the Cygnus Cocoon measured by Fermi-LAT \cite{Ackermann:2011}. The proton injection luminosity is $L_p= 10^{37}\rm erg/s$ with the acceleration spectrum  $E_p^{-2.25}\exp(-E_p/5\,\rm PeV)$. The diffusion coefficient is $D(E_p)=3\times 10^{26}(E_p/1\rm TeV)^{0.7} ~\rm cm^2s^{-1}$. 
    The black solid, blue dashed, and red dot-dashed curves showcase the emission of the entire $6^\circ$ bubble (including the DGE from this region), the emission from interactions between injected protons and MCs, the emission from interactions between protons and atomic gas, respectively.  Panels B, C, D show the measured surface brightness profile in the energy ranges of $2-20$\,TeV, $25-100$\,TeV and $>100\,$TeV (red crosses), in comparison with the model prediction (black curves). Dotted curves show the expected contribution of GDE. See Supplementary Material for details of the model.}
    \label{fig:model}
\end{figure*}

\begin{figure}
    \centering
    \includegraphics[width=0.5\textwidth]{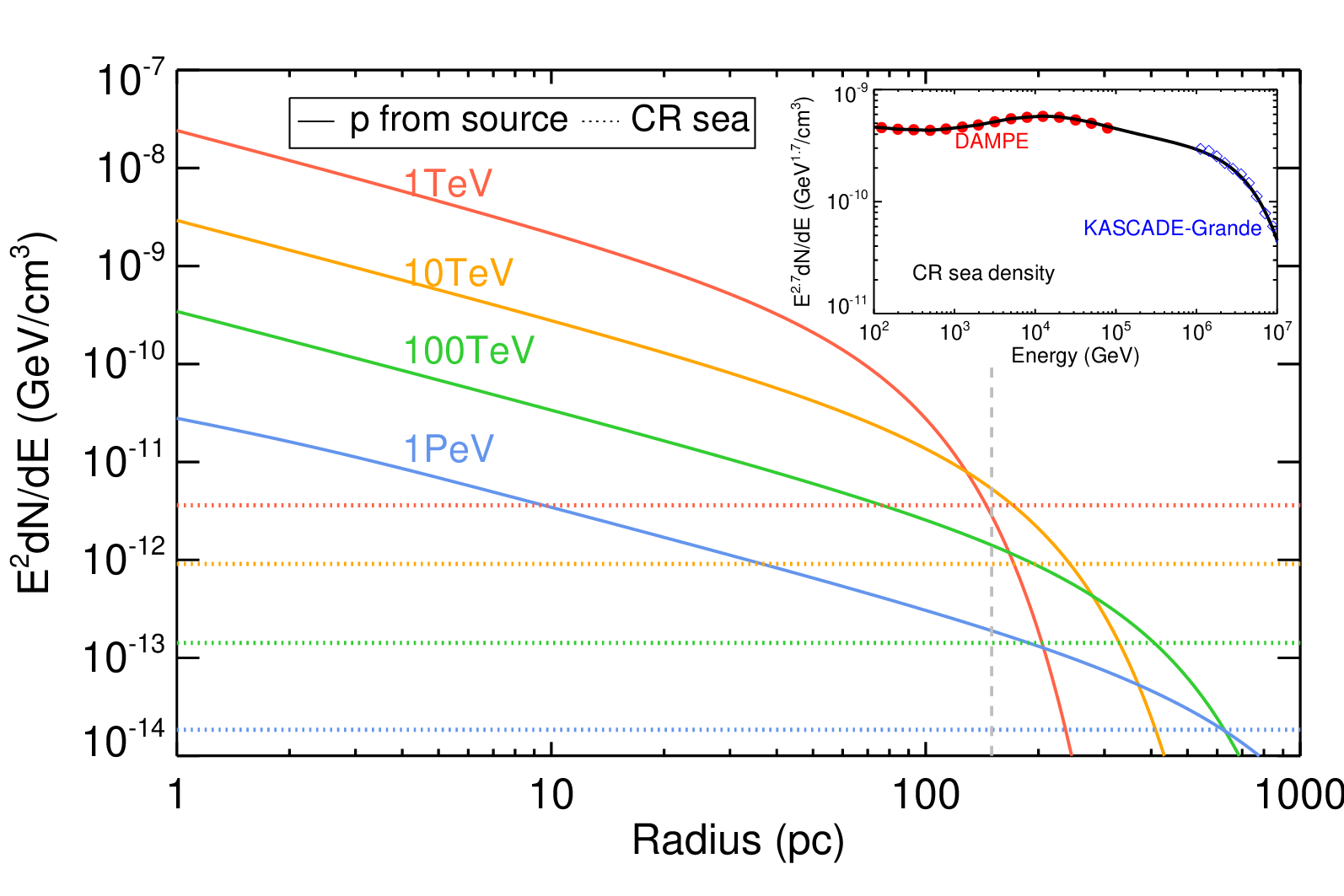}
    \caption{Energy density of relativistic protons as a function of radius from the centre. Solid curves show the energy densities of injected protons, whereas dotted curves mark the energy density of CR `sea' represented by the locally measured CRs. Different colors represent different particle energies as labelled. The vertical grey dashed line marks the radius of the bubble measured by LHAASO. The inset figure shows the spectrum for the CR sea based on the measurements by DAMPE \cite{DAMPE2019_CRp} and KASCADE-Grande \cite{KASCADE2013_CRp} for the proton component.}
    \label{fig:model_CRdensity}
\end{figure}

The 3D distributions of the amorphous HI and the clumpy molecular gas components, combined with the spatial distribution of $\gamma$-rays, tell us that relativistic protons should have a relatively steep, close to $\propto 1/r$ type radial distribution towards the center, implying a continuous regime of particle injection. We adopt a homogeneous energy-dependent diffusion coefficient inside the bubble in the form $D=D_0(E_p/1 \, \rm TeV)^\beta$ with $E_p$ being the particle energy, $D_0$ the diffusion coefficient normalized at 1\,TeV and $\beta$ being a constant related to the property of the turbulence in the environment. The initial (acceleration) spectrum of CRs $\dot{Q}_{\rm p} $ is assumed to be $A E_p^{-s}\exp(-E_p/E_0)$, where $E_0$ represents the cutoff energy of the spectrum. The normalization factor $A$ is determined by the total proton injection luminosity $L_{\rm p}$.  For the given density of the ambient gas, the absolute $\gamma$-ray flux is proportional to the CR density $w_{\rm p}$ which depends only on the injection rate and the diffusion coefficient, namely $w_{\rm p} \propto L_{\rm p}/D_0$.  While the proton density $w_p$ is derived directly from the $\gamma$-ray observations and gas distributions, $L_{\rm p}$  is limited by the available kinetic power of the accelerator.  For example, for 
$D_0\simeq 3\times 10^{26} \, \rm cm^2/s$ with $\beta=0.7$,  
the calculations shown in Figure~\ref{fig:model}, require 
$L_p\simeq 10^{37}\,\rm erg/s$, which,  is about 1\% of the kinetic power of the collective stellar winds in the case of Cygnus OB2.  The diffusion coefficient is about $2-3$ orders of magnitude smaller than the standard value of the diffusion coefficient in the interstellar medium (ISM) derived from the observations of secondary cosmic rays \cite{strong07}. For any reasonable acceleration mechanism, the fraction of the wind mechanical energy converted to CR energy can hardly exceed 10\%. Thus, it is unlikely to increase $D_0$ by more than one order of magnitude, implying that in any realistic model, the CR diffusion inside the bubble should be much slower than that in ISM. The model predicts significant excess of $\geq 100 \, \rm TeV$ proton density (compared to the CR sea)  up to several hundred parsecs  (see  Figure~\ref{fig:model_CRdensity}).  Therefore, this scenario predicts the bubble extension in UHE $\gamma$-rays out of $10^\circ$.  However, because of the low brightness, detecting the outskirts of the bubble is a difficult task. 


The proposed model roughly reproduces the 1D $\gamma$-ray intensity profiles along with the SED of the bubble, as demonstrated in Figure~\ref{fig:model}. Note that the exponential cutoff energy $E_0$ is a formal fitting parameter and does not represent the end of the injection spectrum. The spectrum extends to at least 10\,PeV as demanded by the detection of photons above 1\,PeV, although $E_0=5\,$PeV is employed in the present model. In the model, the $2^\circ$ Gaussian component is considered to be of the same origin as the Cygnus Cocoon measured by Fermi-LAT in the GeV band, which can be ascribed to the interactions between injected CRs and atomic gas within 150\,pc from the center. The hot spots coincident with molecular clouds are explained as well if the molecular clouds are located at $\sim 100\,$pc from the center with an inclination angle of $30^\circ$ with respect to the observer's line of sight, so that the projection distance between the hot spots and the center is about 50\,pc (or $2^\circ$). 

Note that the compact Gaussian component LHAASO~J2031+4057 and the extended Gaussian component LHAASO~J2027+4119 are not necessarily two independent sources. The steep decline in the 1D intensity profile at small radius, made by the superposition of the two Gaussian components, can be more or less reproduced by the $1/r$-type distribution of injected CRs after the diffusion. On the other hand, however, we observe the excess of flux in the observational data with respect to our model prediction at the core region. It implies a larger product of the CR and gas densities in the core region to fit the spectrum of the core simultaneously with the entire bubble, or simply suggests existence of additional UHE $\gamma$-ray sources in the region. In particular, at energy above 400\,TeV, as seen from Figure \ref{Fig:400TeV-bubble}, 7 of the 66 photons are located inside a compact region with a radius of about $0.5^{\circ}$. For PeV photons, 2 of 8 also fall in this rather small circle. For a rough estimate, assuming $1/r$ radial distribution of CRs and a constant gas density in the entire region and taking into account the projection effect \citep{Yang22}, the number of photons within $0.5^{\circ}$ should contribute only about $2\%$ to the total number of photons in the 6-degree bubble, corresponding to about $1.3$ for photons above $400~\rm TeV$ and $0.2$ above $1~\rm PeV$. This excess in the core region by the factor of 5-10  could be caused by more effective confinement of CRs inside Cygnus OB2, given the high turbulent conditions therein. On the other hand, we cannot exclude the possibility that the excess of $\gamma$-ray photons in the core region is related to other independent sources such as the microquasar Cyg X-3 and the pulsar PSR~J2032+4127 or other unresolved $\gamma$-ray emitters. Another possibility is the ballistic propagation effect, which may operate within the distance $r<D/c$ \citep{prosekin15} from the injection center. In this case, the flux at such high energies from the core region could be enhanced due to the strong relativistic beaming effects. Nevertheless, the uncertainty in the core region would not influence the overall fitting of the emission from the entire bubble, as the contribution from the core region to the entire bubble is subdominant.

The high precision of $gamma$-ray spectral measurements over three decades, as shown in Figure \ref{Fig:cen-bubble-SED},  allow robust model-independent predictions of the accompanying neutrino fluxes from 1 TeV to 1 PeV from the Cygnus Bubble (including that from CR sea): $1.1\times 10^{-11}\,\rm cm^{-2}s^{-1}$ above 1\,TeV, $1.9\times 10^{-13}\,\rm cm^{-2}s^{-1}$ above 10\,TeV and $2.1\times 10^{-15}\, \rm cm^{-2}s^{-1}$ above 100\,TeV.

\section{Conclusions and Discussions}
In this letter, we report the detection of an enormous \gray bubble toward the massive star-forming region Cygnus X in the energy rang extending to few PeV. The LHAASO measurements of the \gray flux reveal  the angular size of the structure to at least $6^\circ$ in radius. The spectrum of the bubble is found following a logparabola form over the energy range from 2 TeV to 2 PeV. The detected 8 photons above 1 PeV in the spectrum spread over the entire bubble region make unlikely the leptonic origin of the UHE radiation and indicate 
the presence of  Super PeVatron(s) accelerating protons to energies beyond 10~PeV. 
The inner part of this bubble has its counterpart at GeV energies  
as the Cygnus Cocoon discovered earlier by Fermi-LAT \citep{Ackermann:2011} and by ARGO-YBJ \citep{Bartoli:2014} and HAWC \citep{HAWC-Cocoon} at TeV energies.

The joint analysis of the 
angular distribution of $\gamma$-rays and the atomic hydrogen and CO maps, two objects from totally independent measurements associated together at certain level, allows us to extract information about the radial distribution of protons. It is close to the classical $1/r$ type distribution, implying that we deal with a super PeVatron(s) located within $\sim 0.5^\circ$ region at the center of the bubble,  corresponding to the 15~pc core of the bubble. This region harbors several unidentified multi-TeV sources, one of which has been discovered in this study. 

Taking into account the association with the gases in the Local Arm \citep{xu16}, the core is considered spatially coincident with the massive OB association Cygnus OB2. This location, as well as the age ($\sim 10^7$~yr),  
the speed of stellar winds ($\sim 3000 \ \rm km/s$, and their 
collective mechanical power  ($\sim 10^{39} \ \rm erg/s$) make this object a perfect candidate to act as a CR factory injecting GeV to multi-PeV protons into the circumstellar medium and, in this way,  powering the UHE $\gamma$-ray bubble. The best fit of the spatial and spectral distributions of $\gamma$-rays  is achieved assuming a hard, $E^{-2.2}$ type, proton acceleration spectrum with an exponential cutoff around 5\,PeV. Given the maximum mechanical power contained in the stellar winds of Cygnus OB2,  the transport speed of protons diffusively propagating in the bubble is found to be at least two orders of magnitude slower than in the interstellar medium to explain the observed $\gamma$-ray fluxes. 

The Cygnus Bubble provides us a good example of concentration of CRs surrounding UHE $\gamma$-ray sources. A robust assumption is that every CR source in our Galaxy has such a bubble or halo of CRs. The superposition of the fluxes from the bubbles/halos, particularly many of dim unresolved ones, along the line of sight could make considerable contribution to GDE, as the recent LHAASO measurements that significantly exceeds the theoretical predictions based on conservative estimates of fluxes caused by interactions of CRs with the interstellar gas.

\bibliographystyle{elsarticle-num} 
\bibliography{scibib}

\section*{Acknowledgements}
This work is supported in China by National Key R\&D programme of China 
under grants 2018YFA0404201, 2018YFA0404202, 2018YFA0404203 and 2018YFA0404204 
and by NSFC (grant numbers 12105294,12022502,12261160362,12205314,U1931201), and in Thailand by the National Science and Technology Development Agency (NSTDA) and National Research Council of Thailand (NRCT): High-Potential Research Team Grant Program (N42A650868). We thank all staff members who work year-round at the LHAASO site at 4,400 m above sea level to maintain the detector and keep the electricity power supply and other components of the experiment operating smoothly. We are grateful to Chengdu Management Committee of Tianfu New Area for constant financial support to research with LHAASO data.

\section*{Author Contributions}
C. Li is the convener of this project with proposing and leading the data analysis. Zhen Cao is the spokesperson of the collaboration and coordinated the specific working group involving corresponding authors and relevant members for the project.  F. Aharonian and Zhen Cao prepared the manuscript. S.Z. Chen  optimized the method of cosmic ray background estimation for the analysis.  C. Li and S.Z. Chen analyzed the KM2A data, and S.Q. Xi(supervised by S.Z. Chen) cross-checked the result. M. Zha and C.D. Gao analysed the WCDA data, and R.Y. Tang (supervised by H. Zhou)  cross-checked the result. S.Q. Xi  and S.C. Hu (supervised by M. Zha) estimated the galactic diffuse $gamma$-ray emission for KM2A and WCDA. R.Y. Liu, R.Z. Yang and J. Li analyzed the multi-wavelength data, including the gas distribution analysis, and performed the phenomenological interpretation. The whole LHAASO collaboration contributed to the publication, with involvement at various stages ranging from the design, construction and operation of the instrument, to the development and maintenance of all software for data calibration, data reconstruction and data analysis. All authors reviewed, discussed and commented on the present results and on the manuscript.
\newpage
\onecolumn

\centerline{\Large \bf LHAASO~Collaboration}

\noindent 
Zhen Cao$^{1,2,3}$,
F. Aharonian$^{4,5}$,
Q. An$^{6,7}$,
Axikegu$^{8}$,
Y.X. Bai$^{1,3}$,
Y.W. Bao$^{9}$,
D. Bastieri$^{10}$,
X.J. Bi$^{1,2,3}$,
Y.J. Bi$^{1,3}$,
J.T. Cai$^{10}$
% ...
\\
$^{1}$ Key Laboratory of Particle Astrophyics \& Experimental Physics Division \& Computing Center, Institute of High Energy Physics, Chinese Academy of Sciences, 100049 Beijing, China\\
$^{2}$ University of Chinese Academy of Sciences, 100049 Beijing, China\\
$^{3}$ TIANFU Cosmic Ray Research Center, Chengdu, Sichuan,  China\\
$^{4}$ Dublin Institute for Advanced Studies, 31 Fitzwilliam Place, 2 Dublin, Ireland \\
$^{5}$ Max-Planck-Institut for Nuclear Physics, P.O. Box 103980, 69029  Heidelberg, Germany\\
$^{6}$ State Key Laboratory of Particle Detection and Electronics, China\\
$^{7}$ University of Science and Technology of China, 230026 Hefei, Anhui, China\\
$^{8}$ School of Physical Science and Technology \&  School of Information Science and Technology, Southwest Jiaotong University, 610031 Chengdu, Sichuan, China\\
$^{9}$ School of Astronomy and Space Science, Nanjing University, 210023 Nanjing, Jiangsu, China\\
$^{10}$ Center for Astrophysics, Guangzhou University, 510006 Guangzhou, Guangdong, China\\
% ...




\onecolumn
\section*{Supplementary materials}
\section{Analysis of LHAASO data}
\label{sec:analysis}
\subsection{Data}
The low energy $\gamma$-ray data below 10 TeV are from LHAASO-WCDA since March 5th, 2021 when the whole detector of 78,000 m$^2$ has been fully operated. By May 31th, 2023, in total the exposure time is 6.35$\times10^{7}$ seconds for air shower event collection. The trigger algorithm for an event requires at least $\ge$30  {\it hits} appearing in any cluster of 12$\times$12 detector units within a time window of 250 ns, where a {\it hit} is formed in a detector unit once a 8-inch photo multiplier tube (PMT) received a signal more than 1/3 photo electrons (PE) or a 20-inch PMT received a signal more than 1 PE within 100 ns. The average event rate is 35 kHz. Events with the arriving zenith angle $\le$50$^\circ$ are used in the analysis to maintain the high event reconstruction accuracy. In a surveying cycle of 24 hours for the entire northern sky, 1.29$\times10^9$ $\gamma$-like events are selected by applying the $\gamma/hadron$ discrimination criterion on the air shower events. The criterion requires the air shower lateral distribution to be smooth enough, as the characteristic of a $\gamma$-ray induced air shower, namely the total relative deviation  $\Sigma_i(\Delta \zeta_i /\sigma)^2$ over all hits in an event must be smaller than a certain value, $P_m$. Here, $\zeta_i$ is the logarithmic of the measured charge $Q_i$ of $i$-th hit,  $log_{10}(Q_i)$. The deviation $\Delta \zeta_i$ is defined as the difference between the measured $\zeta_i$ and the corresponding value of an ideally smooth lateral distribution that is an average over many $\gamma$-like events selected from the Crab direction with similar energies. The deviation must be a function of the lateral distance of the hit from the shower core, namely larger deviations are expected at further positions where the charge drops to much smaller numbers compared with the closer ones, so they should be normalized by an averaged fluctuation $\sigma$, which is also the function of the distance from the shower core, before they are summed over all hits in the shower. The averaged fluctuation, $\sigma$ and $P_m$ are determined by optimizing the significance of the measurement of the standard candle, the Crab Nebula. As a test, the  Crab is observed with a significance of 328.11$\sigma$ using the current data set. The SED is also found to be  the same as reported previously\cite{2021Sci...373..425L}. 

The high energy $\gamma$-ray data above 10 TeV are taken using LHAASO-KM2A since December 2019 when the first half-array was operated, using 3/4 array starting from December 2020, and the full array starting from July 2021. 
The array is triggered once more than 20 EDs are fired within a time window of 400 ns. Then all hits within $5\,{\mu}$s before and after the trigger time will be recorded for off-line analysis. With the calibration of each detector, the ADC counts of EDs and MDs are converted into numbers of particles. The status of each detector is monitored in real time, and only detectors under normal conditions are used in the shower reconstruction. Both measured and simulated events are processed through the same reconstruction pipeline. The direction of showers is reconstructed by fitting the relative arriving time of the ED hits. The angular resolution (denoted as $\phi_{68}$, containing $68\%$ of the events) is $0.24^{\circ}$ at 100 TeV for $gamma$-ray showers with zenith angles less than $35^{\circ}$. The parameter $\rho_{50}$, defined as the particle density at 50 m from the shower axis obtained by fitting the modified NKG function to hits in a shower, is used as the energy estimator. The energy resolution is about $13\%$ at 100 TeV for showers with zenith angle less than $20^{\circ}$. The ratio between the measured numbers of muons and electrons in a shower is used to discriminate  electromagnetic showers from hadronic showers. The rejection power is about $4\times10^3$ at energies above 100 TeV after discrimination cuts for half array, and is better at higher energies. The details about the calibration and shower reconstruction and selection are described in \cite{2021ChPhC..45b5002A}. In order to achieve a better background rejection power for extended source analysis, the ratio between muons and electrons used in this analysis is stricter than that in previous point source analysis and only events with zenith angle less than 40$^{\circ}$ are used. The cosmic ray survival ratio is estimated to be less than 1$\times10^{-4}$ above 100TeV with this tight cut for full array, while the survival ratio of gamma rays is estimated to be 70\%.

\subsection{3-D likelihood analysis}
Considering the very complex nature of the Cygnus X, a 3-dimensional likelihood fitting method is developed in this analysis. Data at different energy bins are fitted simultaneously by assuming a spatial template and an energy spectrum form for each source in the region of interest. The point spread function (PSF) as a function of energy is convoluted with the spatial template. The detector responses as a function of energy for different arrays are considered according to the detailed detector simulation. In this way, both morphological and spectral information are taken into consideration to account for contributions from individual  sources.

The statistic quantity used to evaluate the significance of the test is TS $= 2 \log (\lambda)$, where $\lambda={\mathcal{L}_{s+b}}/{\mathcal{L}_{b}}$. ${\mathcal{L}_{s+b}}$ is the maximum likelihood value for the source plus background hypothesis, while $\mathcal{L}_{b}$ is for the background-only hypothesis. The combination of all sources including target source and gas templates is regarded as an alternative hypothesis, and all other sources(including gas templates) are considered as a null hypothesis. The spectral parameters are fitted in the alternative hypothesis. According to Wilks' Theorem\cite{wilks1938}, TS follows a chi-square distribution with the number of degrees of freedom equal to the difference in the number of free parameters between the hypotheses.  

The map of the TS values obtained in the 3D-likelihood fitting procedure is used to represent the morphology of the region of sources. Particularly in the analysis, a 2-dimensional Gaussian template with a spectral index of -2.6 for WCDA and -3.0 for KM2A in every pixel of the map is assumed. The spectral indexes are chosen to present the most common situations for sources measured by the two detector systems. In this way, the TS map of each source is separated out individually according to the best-fit model.

\subsection{Fitting results}
The region of interest (ROI) in the analysis is a disk with a radius of 10$^{\circ}$ centered at LHAASO J2032+4102. A circular area with a radius of 2.5$^{\circ}$ around the very bright unidentified source, LHAASO J2018+3651, are masked. The radius is chosen to ensure that almost all of the signals from LHAASO J2018+3651 and nearby sources are masked even considering the PSF of detectors. Taking the results of previous investigations as guidance, the $\gamma$-ray emission in this region is assumed to be due to three sources as the initial trial, namely the unidentified source at the center, $\gamma$-Cygni , and an extended source. Clearly, more features in the LHAASO data for the Cygnus X region than the assumption have been uncovered. First of all, hot spots coincided with MCs with significance exceed 5$\sigma$ are still left after all three sources removed. Secondly,  several new sources are revealed at the Cygnus direction. 
Thirdly, there is a more extended structure which has a much bigger extension than what learned before as the cocoon of $\sim2^\circ$ and follows the atom distribution. The distribution of atom(HI) and molecular clouds(MC) in the line of sight are used as templates to fit the data. 
The fitting results for HI and MC templates as well as the Gaussian component are displayed at Table.1. 
The HI and MC templates are  detected with TS values of 108 and 88, separately by KM2A. 
The details about the atom and molecular clouds distributions will be described in the following sections.

Even with gas distribution templates added, there are still some spots with significance exceeding 5$\sigma$, which indicates there are other sources located within the ROI region. The method mentioned above is used to test whether there is another source. New sources are added in the fitting until the $\Delta$TS is less than 25. 
In total, there are 11 sources located at the Cygnus Bubble region. We performed a global fitting using 11 sources plus the gas templates simultaneously. Most of the sources are located at the peripheral region, and the flux is only about few percents of the bubble. 

\begin{figure*}[htbp]
  \centering
  \begin{tikzpicture}
  
    \node at (0+2.6,0-0.17) {\includegraphics[width=0.373\textwidth]{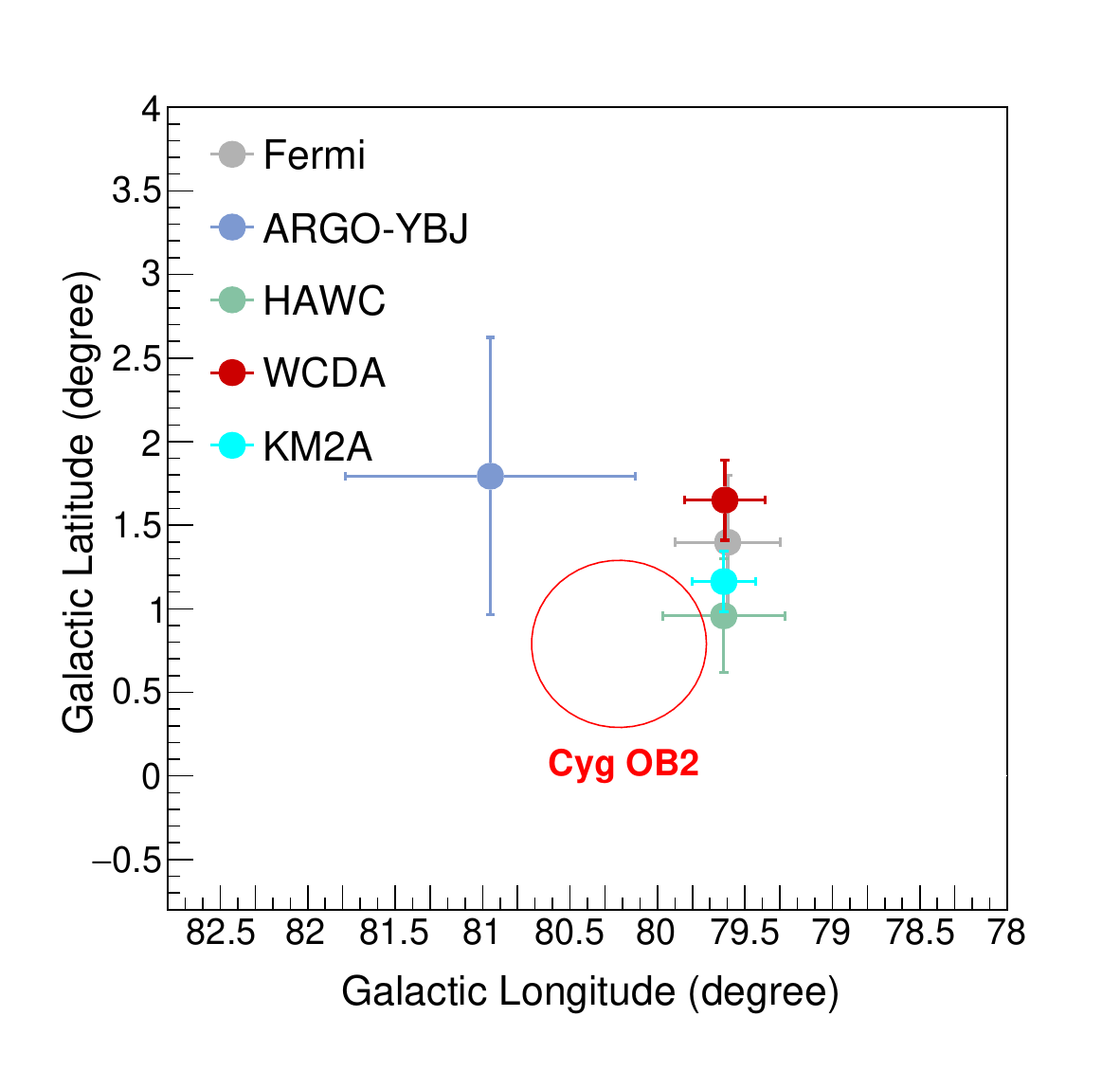}};
    \node at (9+2.3,0) {\includegraphics[width=0.63\textwidth]{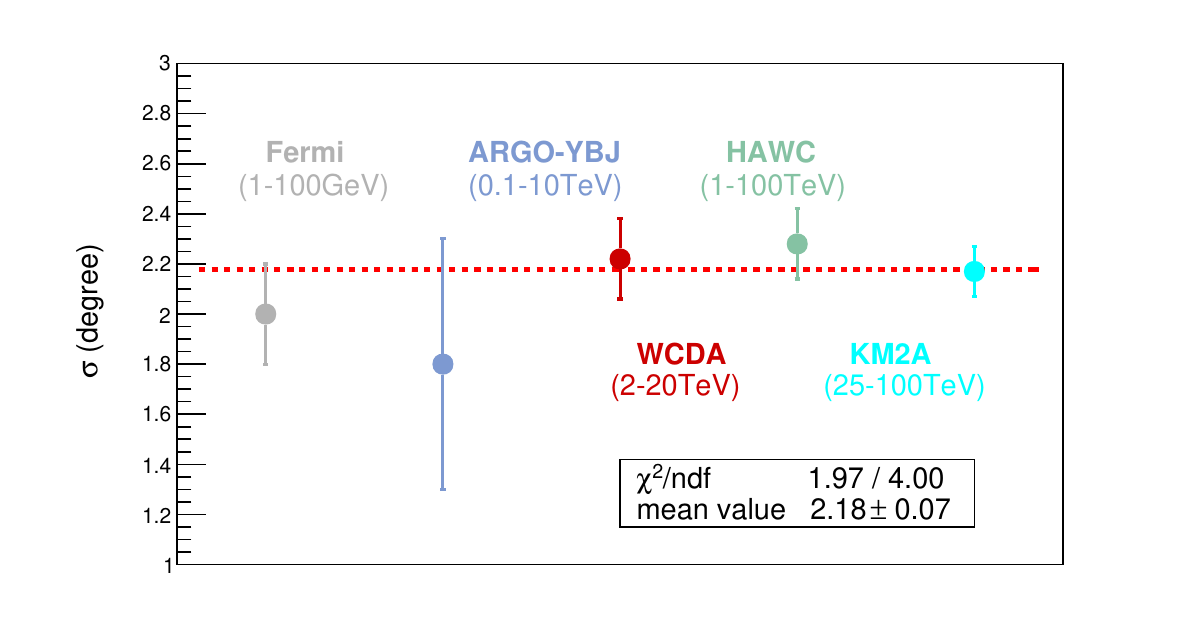}};
    
  \end{tikzpicture}
  \caption{Left: The positions of the Cygnus Bubble measured by LHAASO KM2A and WCDA as the center of the inner bubble, other results are from Fermi-LAT  (dubbed it Cocoon\cite{Ackermann:2011}), HAWC\cite{HAWC-Cocoon} and ARGO-YBJ\cite{Bartoli:2014}. 
  A region with a radius of 0.5$^\circ$ relative to the center of Cygnus OB2 (red circle) is plotted here for reference. 
  Right: The extensions of the inner bubble measured by the experiments. No significant variation is observed over energies from 1 GeV to around 1 PeV.}
  \label{Cocoon-pos}
\end{figure*}

A very extended source with a Gaussian width of $2.17^\circ\pm0.10^\circ$ at the best-fit position of right ascension (RA) = $307.43\pm0.16^\circ$ and declination (DEC) = $41.05^\circ\pm0.13^\circ$ is detected by KM2A
named as LHAASO J2027+4119. 
The normalization factor $N_{0} = (0.62 \pm 0.05)\times10^{-11} \rm m^{-2}TeV^{-1}s^{-1}$ and the spectral index $\Gamma = -2.99\pm 0.07$. No clear indication of cut-off of the SED is found.
This source is also well measured in the energy range below 20 TeV using LHAASO-WCDA. The extension is found to be $2.28^\circ\pm 0.14^\circ$  as the Gaussian width. The best-fit position is RA = $306.90^\circ\pm0.23^\circ$ and DEC = $41.33^\circ\pm0.16^\circ$. The SED is described in the form of $dN/dE = N_{0}(E/E{_{0}})^\Gamma$, where the normalization factor is $N_{0} = (1.27\pm 0.14)\times 10^{-9} \rm m^{-2}TeV^{-1}s^{-1}$ at $E_{0}$ = 7 TeV and the spectrum index is $\Gamma = -2.63\pm0.08$. 
The position and extension of this source from other experiments are also shown in Figure \ref{Cocoon-pos}. The position is well consistent with the results from Fermi and HAWC experiments, and within 2$\sigma$ of ARGO-YBJ result. The mean extension is around 2.18$^{\circ}$, which shows a sign of larger extension above tens of TeV, however the significance is rather low.

The core region is very complex,  which contains Cygnus OB2, the unidentified source LHAASO J2031+4127 and the microquasar Cygnus X3 and so on. 
The current LHAASO data analysis reveals that there are 3 sources resolved in this region. 
Two of them (LHAASO J2031+4141 and LHAASO J2032+4125) are positionally close to the TeV source LHAASO J2031+4127 and both reveal a 
spectral cutoff around $30~\rm TeV$, which is significantly different from the spectral behavior of Cygnus Bubble. The third one is a new source (named LHAASO J2031+4057) detected by WCDA, which has a similar spectrum index to the bubble.  
It is modeled using a 2D Gaussian spatial template.  The Gaussian width is $0.33 \pm 0.08^{\circ}$, and the best-fit position is RA=$307.89 \pm 0.09 ^{\circ}$, DEC=$40.96 \pm 0.16 ^{\circ}$. The spectrum is described in the form of $f(E)=N_{0} (\frac{E}{E_{0}})^{\Gamma}$ , where the normalization factor is $N_{0}=(0.11 \pm 0.06) \times 10^{-9} \rm TeV^{-1} m^{-2} s^{-1}$ at $E_{0}=7$ TeV and the spetrum index is $\Gamma =-2.75 \pm 0.17$. 

The source at RA=$305.17^\circ\pm0.07^\circ$ and DEC=$40.44^\circ\pm0.05^\circ$ (LHAASO J2021+4030) with a Gaussian width of $0.23^{\circ}\pm0.04^{\circ}$ is found to be coincident with the SNR $\gamma$-Cygni. Its SED is described by a single power law functional form. Details about the source are beyond the scope of this paper and will be published elsewhere. The rest sources lie at the edge of ROI, which has tiny influence on Cygnus Bubble.


\begin{table*}
\centering
\resizebox{\linewidth}{!}{
\begin{tabular*}{22cm}{llcccccc}
\hline
\hline
 Source & Components & $\alpha_{2000}(\circ)$ & $\delta_{2000}(\circ)$ & $r_{39}(\circ)$ & TS & $N_{0}(TeV^{-1} m^{-2} s^{-1})$ & $\Gamma$\\
\hline
 LHAASO J2027+4119 & KM2A & $307.43\pm0.16$ & $41.05\pm0.13$ & $2.17\pm0.10$ & 145 & $(0.62\pm0.05)\times10^{-15}@50TeV$  & $-2.99\pm 0.07$ \\
  & WCDA & $306.90\pm0.23$ & $41.33\pm0.16$ & $2.28\pm0.14$ & 251.44 & $(1.27\pm0.14)\times10^{-9}@7TeV$  & $-2.63\pm 0.08$ \\
 \hline
HI & KM2A & ~ & ~ & ~ & 108 & $(0.69\pm0.10)\times10^{-15}@50TeV$ & $-2.94\pm 0.12$  \\
  & WCDA & ~ & ~ & ~ & 60.77 & $(1.43\pm0.26)\times10^{-9}@7TeV$ & $-2.66\pm 0.12$  \\
 \hline
 MC & KM2A & & & & 88 & $(0.46\pm0.06)\times10^{-15}@50TeV$ &  $-2.87\pm 0.14$  \\
 & WCDA & & & & 67.47 & $(1.08\pm0.19)\times10^{-9}@7TeV$ &  $-2.73\pm 0.13$  \\
   \hline
 LHAASO J2031+4057 & WCDA & $307.89\pm0.09$ & $40.96\pm0.16$ & $0.33\pm0.08$ & 115.40 & $(0.11\pm0.06)\times10^{-9}@7TeV$  & $-2.75\pm 0.17$ \\
 
\hline
\hline
\label{Tabel1}
\end{tabular*}
}
\caption{The fitting results for the components of Cygnus Bubble.}
\end{table*}

\subsection{UHE photons}
Benefiting from the excellent performance of KM2A, the CR background is about one order of magnitude lower than the level of signal even at a large region with radius of 6$^\circ$ at energies above hundreds of TeV. The numbers of on-source-events are 14, 6 and 2 for the last three energy bins, respectively. The number of CR-like background events are estimated to be 1.9 ,0.6 and 0.2 correspondingly. Thus we can get valuable information with these individual gamma ray photoes. The detailed information of events with energy above 1\,PeV within 6$^{\circ}$ is listed in Table.\ref{PeVphoton}.

\begin{table*}
\centering
\begin{tabular*}{13cm}{llllllll}
\hline
\hline
 E (PeV) & $\delta$E (PeV) \hspace{0.3cm} & $N_{e}$ \hspace{0.2cm}& $N_{\mu}$ \hspace{0.2cm} &$\theta$($^{\circ}$)& $D_{edge}$($m$) \hspace{0.2cm}& $\psi$($^{\circ}$) \\
\hline
1.08 & 0.16 & 5904 & 13.0 & 19.4 & 143 & 4.7\\
1.19 & 0.18 & 5480 & 14.1 & 34.4 & 73 & 0.2\\
1.20 & 0.18 & 6939 & 12.6 & 14.2 & 132 & 5.8\\
1.35 & 0.20 & 6938 & 8.4 & 27.1 & 43 & 2.9\\
1.38 & 0.20 & 6469 & 8.9 & 17.4 & 52 & 2.6\\
1.42 & 0.21 & 6258 & 6.6 & 12.7 & 57 & 0.1\\
1.78 & 0.27 & 6665 & 12.8  & 18.0 & 41 &1.8\\
2.48 & 0.37 & 13815 & 29.1 & 33.0 & 99 & 5.2\\


\hline
\hline
\end{tabular*}
\caption{Specifications of the photons with energy above 1 PeV. E and $\delta$E are the reconstructed energy and its error. $N_{e}$ and $N_{\mu}$ are the detected number of secondary charged particles and muons. $\theta$ is the incident zenith angle of shower. $D_{edge}$ is the distance of shower core from the nearest edge of array in meters. $\psi$ is the space angle between $\gamma$ like event and the center of ROI}
\label{PeVphoton}
\end{table*}

\subsection{Spectral variations}
The spectral variations across the bubble can help us to understand the propagation of cosmic rays in this region. The brightness of central region overlapped with MC seems higher than other regions, which may indicate the higher cosmic ray flux. To check the spatial variation, We divided the MC template into two parts. The region within $3^\circ$ from the center of the bubble is defined as inner part, while as the region beyond $3^\circ$ is defined as the outer part.  No significant variation of spectral indices is observed between the two parts, while the ratio between the fluxes is about 1:1 between the two parts. According to the gas survey, the ratio of the total gas masses is about 1:2 between the two regions.


\section{Distribution of Molecular Clouds and HI Clouds}
\label{sec:gas}
\begin{figure*}[hbtp]
\centering
\includegraphics[width=0.46\textwidth]{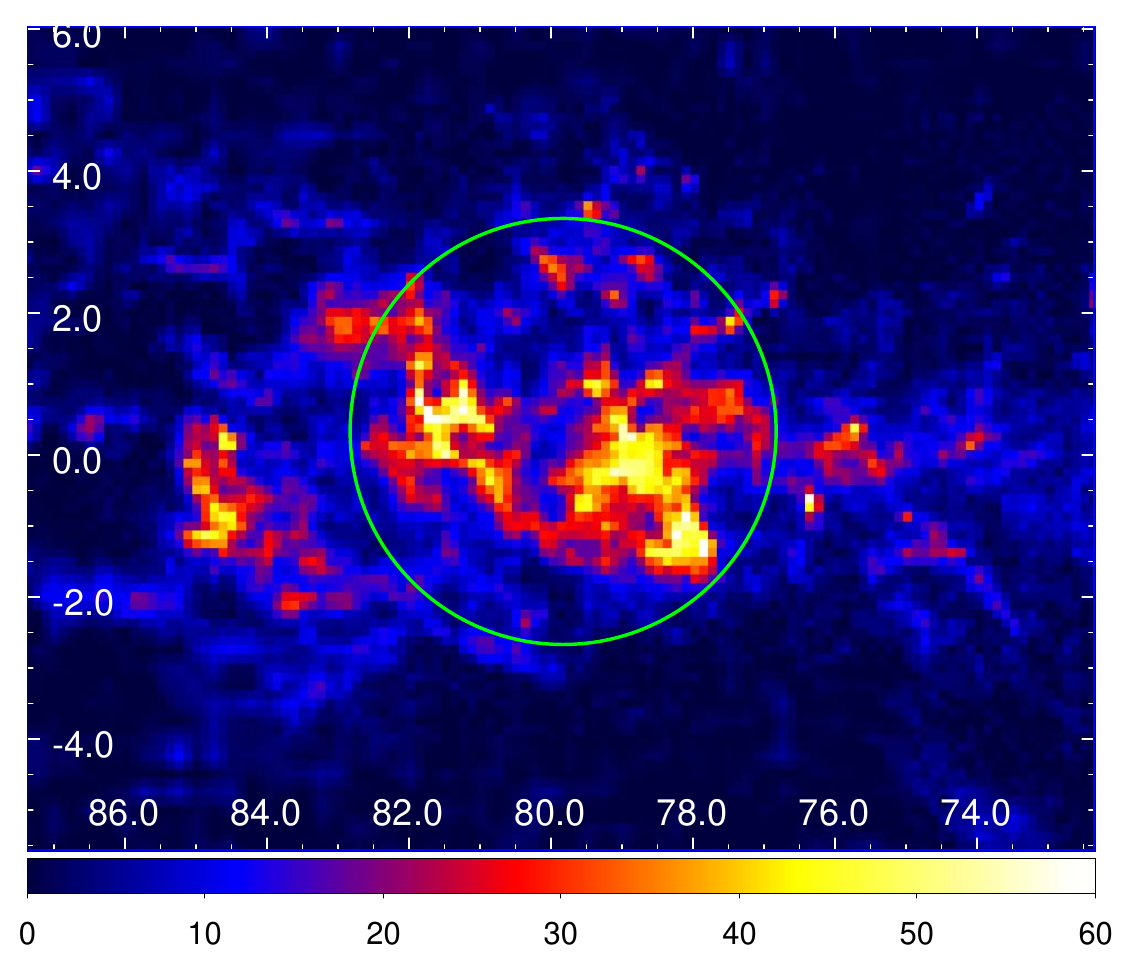}
\includegraphics[width=0.53\textwidth]{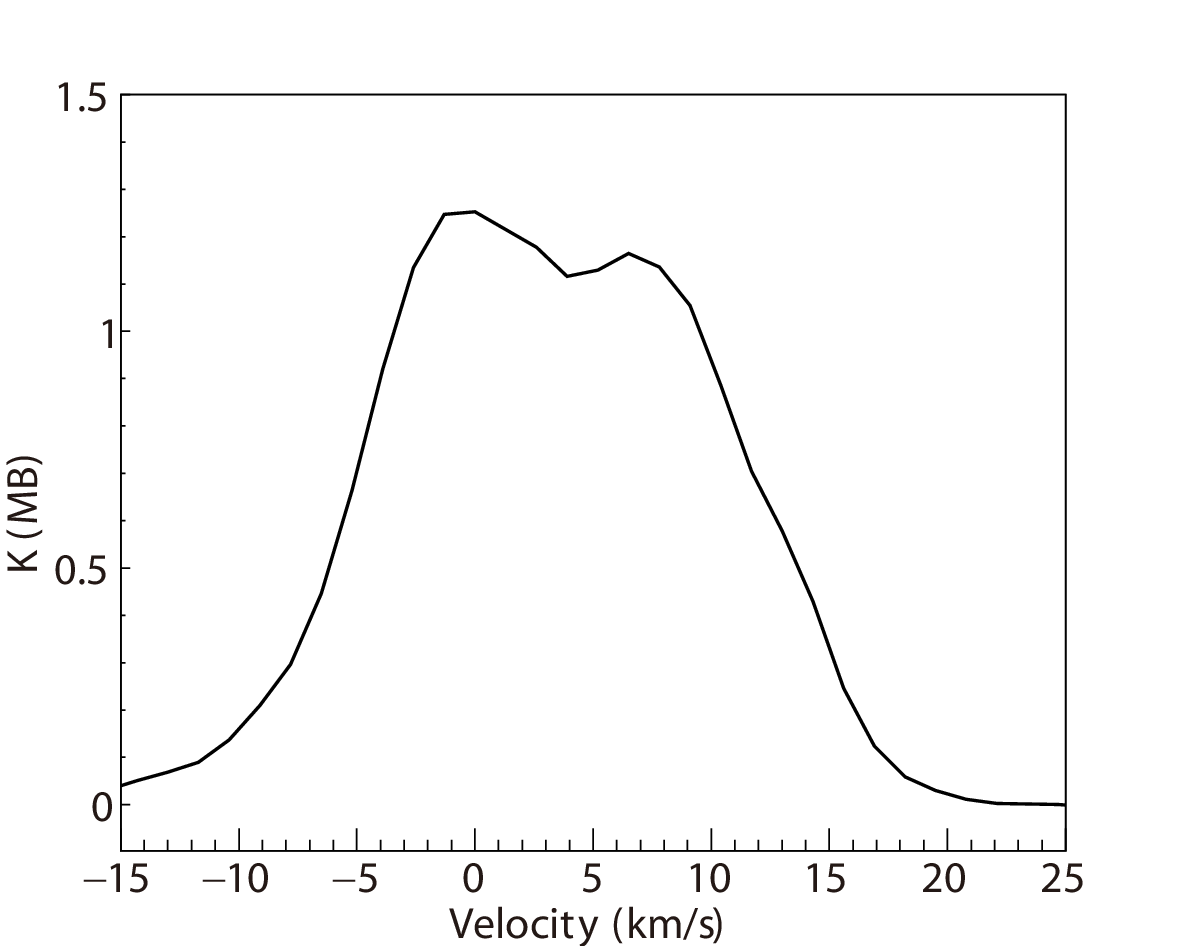}\\
\includegraphics[width=0.46\textwidth]{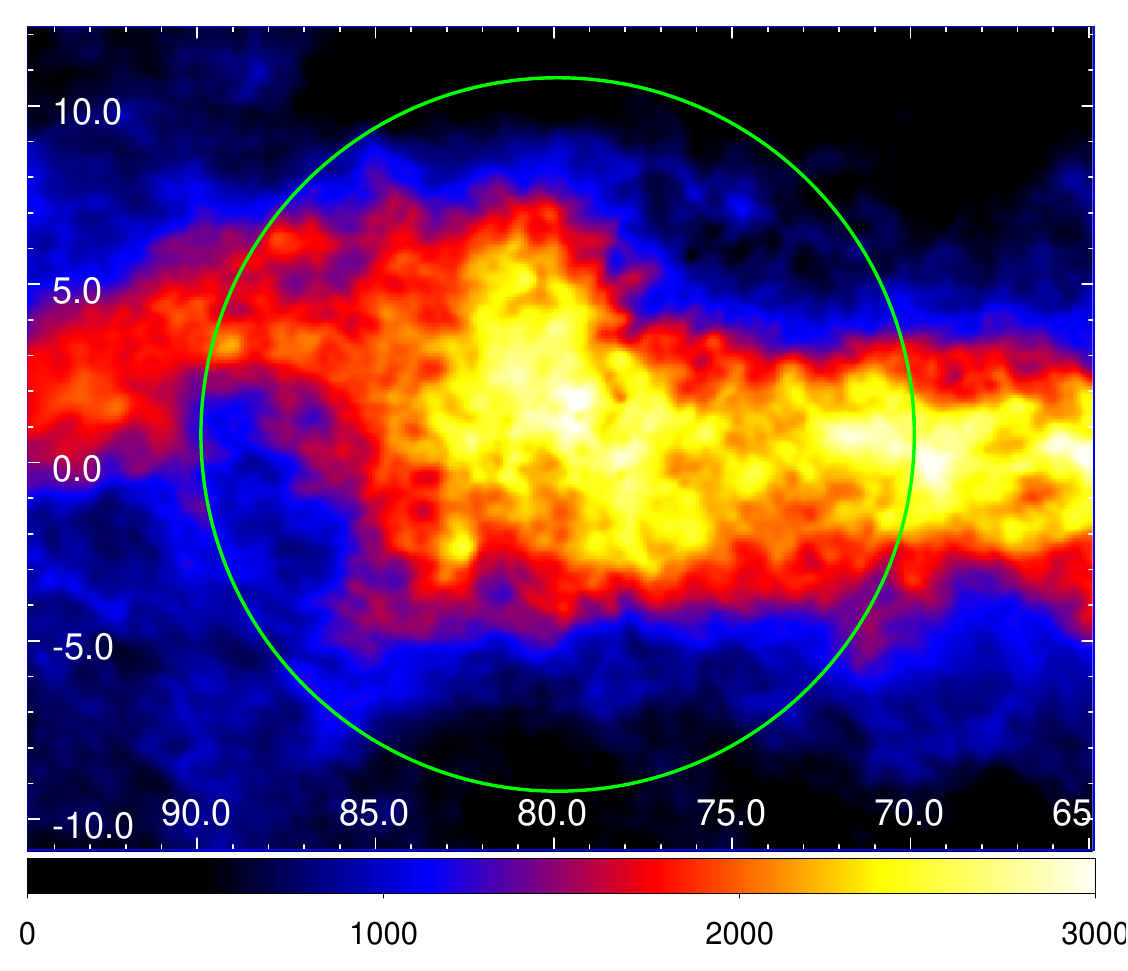}
\includegraphics[width=0.53\textwidth]{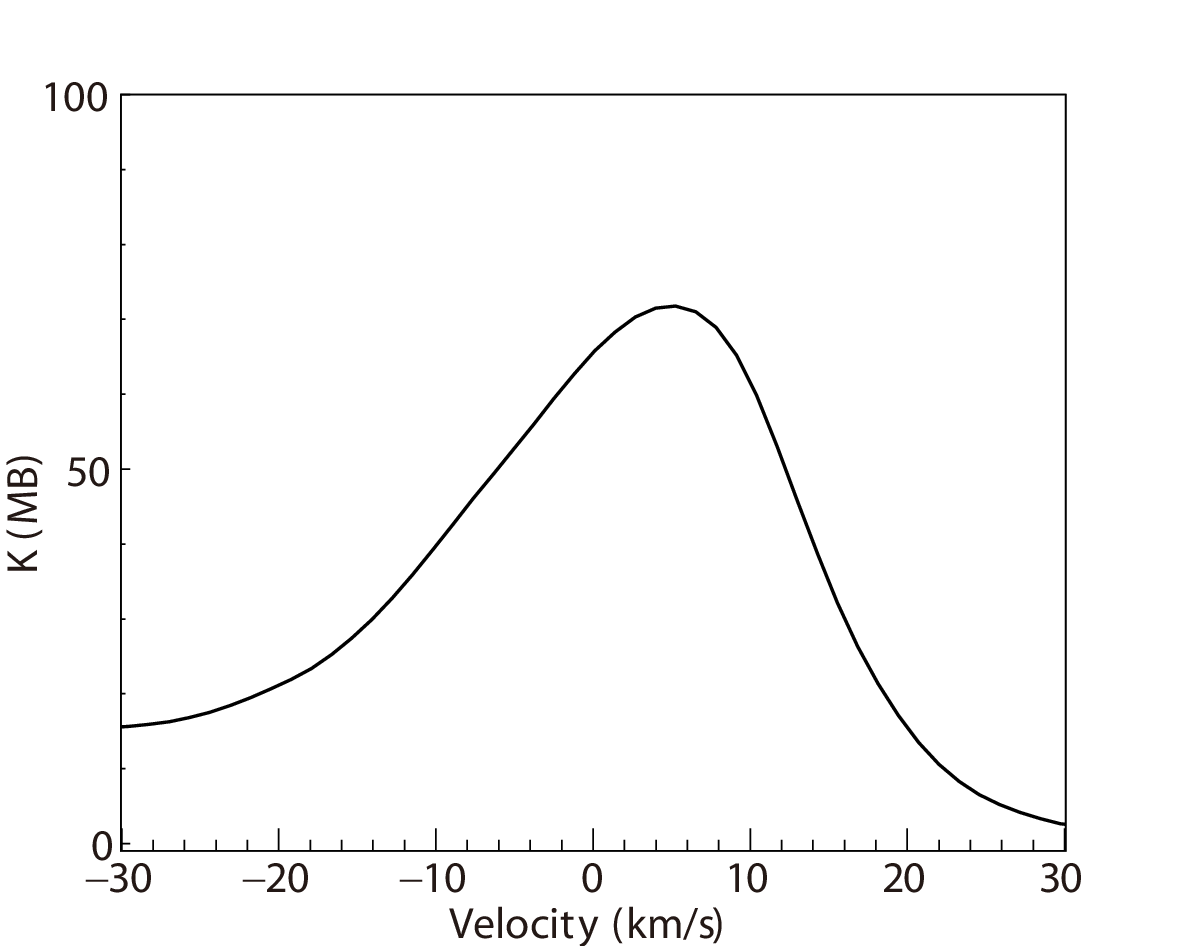}

\caption{Left panels: \twCO (top) and H\,{\sc i} (bottom) intensity maps in Galactic  coordinates (l,b) in degrees integrated over the velocity ranges -10 to $20\km\ps$ and -20 to $30\km\ps$, respectively. 
The color denotes the intensity in unit of K $\km\ps$ 
The regions delineated in green are used to estimate the astrophysical parameters for the molecular gas and atomic gas.
Right panels: \twCO\ and H\,{\sc i} spectra of the regions indicated in left panels.}
\label{CO_HI_MAP_region}
\end{figure*}

We adopted CO data from the CfA 1.2m CO survey\cite{Dame01} in this work.
The \twCO\ (\otz) map integrated over the velocity range -10 $\km\ps$ to 20 $\km\ps$ is shown in Figure \ref{CO_HI_MAP_region} (top left panel), which is associated with the Cygnus OB2 cluster at $\sim$ 1.46 kpc\cite{Mel'nik17}.
Molecular Clouds (MCs) are observed to be spatially coincident with the TeV emissions observed by LHAASO (Figure \ref{Fig:hot-spots-bubbles}).
We estimated the astrophysical properties of the MCs in the centeral regions of LHAASO detected emission (Figure \ref{Fig:hot-spots-bubbles}) and have parameterized the distance as $d=1.46d_{1.46}\kpc$.
The \twCO (\otz) spectrum of the region is extracted and shown in Figure \ref{CO_HI_MAP_region} (top right panel).
The estimated results are summarized in Table~\ref{COparameter}.
The total mass of molecular cloud ($\mbox{H}_2$) within the LHAASO ROI is estimated to be 3.25$\times$ 10$^6 M_\odot$$d_{1.46}^2$

\begin{figure}[hbtp]
\centering
\includegraphics[width=0.5\textwidth]{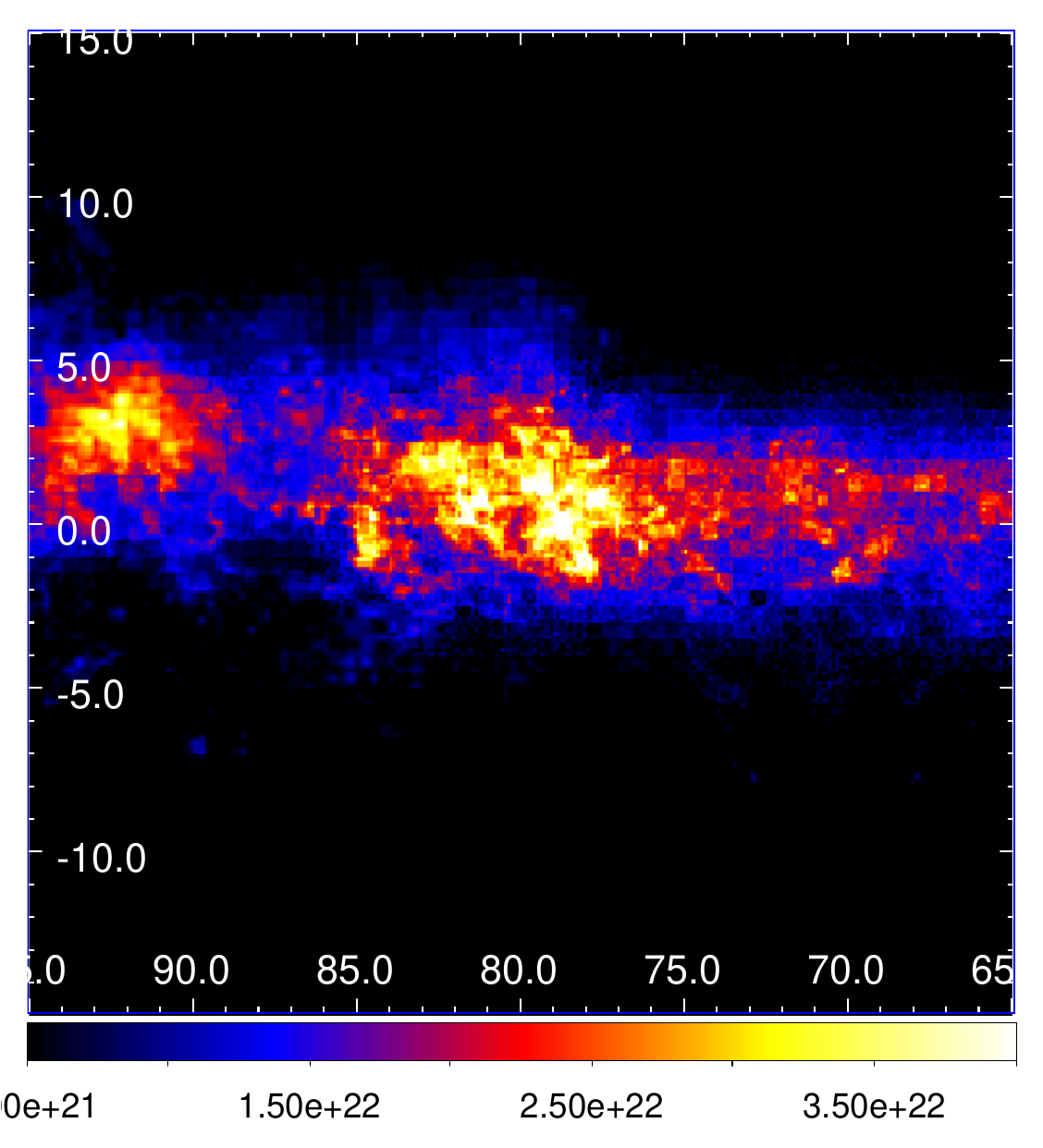}
\caption{The gas template (in units of hydrogen column density) used in the likelihood fitting.}
\label{fig:gastemp}
\end{figure}

To compare the $gamma$-ray emission of the Cygnus Bubble with the large-scale gas distribution in the region, we used the 21 cm emission line of H\,{\sc i} as a tracer of the neutral atomic gas.
The HI 4-PI Survey (HI4PI) survey data\cite{HI4PI} were investigated.
The H\,{\sc i} map for velocity range from -20 $\km\ps$ to 30 $\km\ps$, which corresponds to the $\sim$ 1.46 kpc distance of the Cygnus OB2 cluster\cite{Mel'nik17} is shown in Figure \ref{CO_HI_MAP_region}, bottom left panel.
The green circle indicates the ROI of LHAASO data analysed in this paper.
Large scale diffuse H\,{\sc i} emission is observed to be spatially associated with TeV emission of the Cygnus Bubble (Figure \ref{Fig:hot-spots-bubbles}). 
We estimated the astrophysical properties of the atomic clouds within the LHAASO ROI, and the corresponding spectrum is extracted and shown in Figure \ref{CO_HI_MAP_region} (bottom right panel). 
The estimated results are also summarized in Table~\ref{COparameter}.
We transfer both the CO and H\,{\sc i} emission into hydrogen column density. 
The total hydrogen column density in the direction of Cygnus Bubble is shown in Fig.\ref{fig:gastemp}.

\begin{table*}{}
\centering
\caption{Derived parameters for the molecular and atomic clouds indicated in Figure \ref{CO_HI_MAP_region}}.

\begin{tabular}{cccc}
\\
\\
\hline\hline

 & $N$(H$_2$) & $n(\mbox{H}_2)$  & $M(\mbox{H}_2)$    \\
       & ($10^{21}$cm$^{-2}$)  & (cm$^{-3})$ & $(10^6 M_\odot)$  \\

\hline\hline   
\\
Molecular Cloud & 4 & 17$d_{1.46}^{-1}$ & 1.2$d_{1.46}^2$    \\
\\
\hline\hline

 & $N$(H) & $n(\mbox{H})$  & $M(\mbox{H})$    \\
       & ($10^{21}$cm$^{-2}$)  & (cm$^{-3})$ & $(10^6 M_\odot)$  \\

\hline\hline 
\\
Atomic Cloud & 3.5 & 4.5$d_{1.46}^{-1}$ & 5.9$d_{1.46}^2$    \\
\\

\hline\hline                 
\\
\label{COparameter}
\end{tabular}
\end{table*}

\clearpage 

\section{Detailed description of the modeling and the conclusions} 


Regardless of complex particle acceleration and transport processes in the core source region, we assume relativistic protons are injected into the bubble from a point-like source at a constant rate. 
Considering an isotropic diffusion, the spatial distribution of protons at the present time as a function of the radius from the central source $r$ reads
\begin{equation}\label{eq:crdis}
w_{\rm p}(E_p,r)=\int_0^{t_{\rm age}}dt\frac{\theta(ct-r)Q_{p}(E_p)}{4\pi^{3/2}R_{\rm inj}R_{\rm diff}(E_p,t)r}\exp\left( -\frac{r^2}{R_{\rm diff}^2(E_p,t)}\right)
\end{equation}
where the injection rate $Q_p(E_p)$ is assumed to be a power-law function with a high-energy cutoff, i.e., $Q_p(E_p)=Q_0E_p\exp(-E_p/E_{p, \rm max})$. $Q_0$ is related to the total proton injection luminosity $L_p$ by $L_p=\int E_pQ_P(E_p)dE_p$. $R_{\rm diff}(E_p,t)=\sqrt{4D(E_p)t}$ is the diffusion length of protons with energy $E_p$ injected at a period of time $t$ ago (i.e., $t=0$ means the present time). $\theta(ct-r)$ is the Heaviside function employed here to avoid the unrealistic superluminal diffusion. 


\begin{figure*}
    \centering
    \includegraphics[width=1\textwidth]{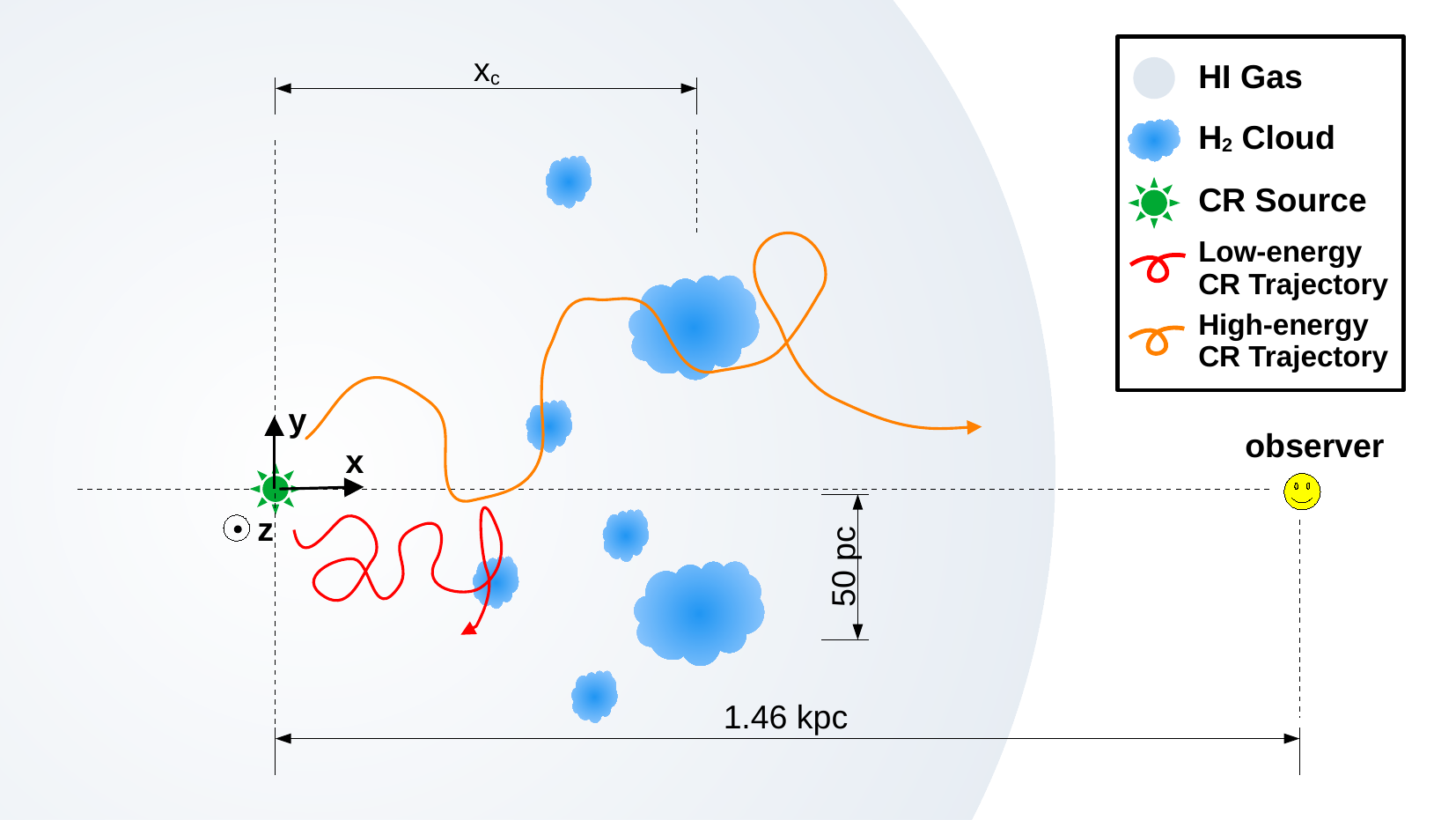}
    \caption{An illustrative sketch (not to scale) of the model.}
    \label{fig:sketch}
\end{figure*}

\begin{figure*}
    \centering
    \includegraphics[width=1\textwidth]{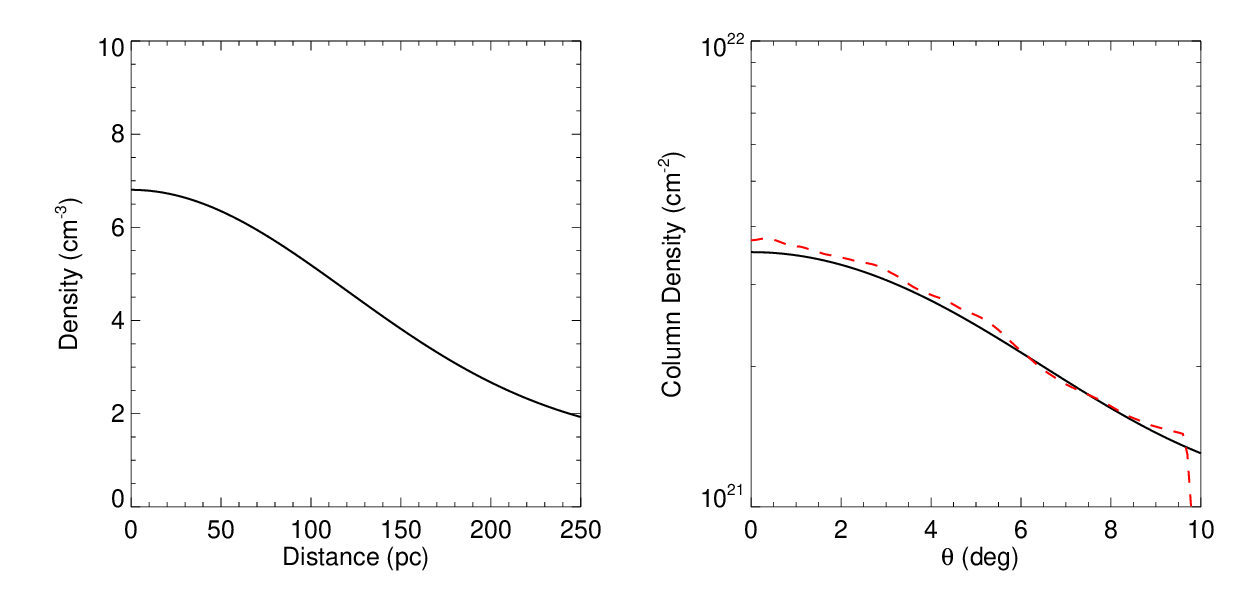}
    \caption{Solid curves show the 1D radial density distribution of HI gas (the left panel) employed in our modeling (see Eq.~\ref{eq:HIdensity}) and the corresponding 1D column density distribution (the right panel). The red dashed curve in the right panel shows the observed 1D radial distribution of the intensity of the 21\,cm HI emission, which is multiplied by an arbitrary factor for comparing the shape of its distribution with the assumed one in modeling.}
    \label{fig:HIgas}
\end{figure*}

While we approximate the particle transport to be isotropic, the gas distribution is clearly not as can be seen from the associations between the hot spots and CO clumps. In fact, the center of the bubble is also slightly deviated away from the central source but we still treat it as an isotropic distribution for simplicity. To approximately reproduce the positional relation between the CR source and the gas distribution, the model should involve the 3D distribution of CRs and gases. A sketch of the physical picture is shown in Fig.~\ref{fig:sketch}. For simplicity, we assume the density distribution of the HI gas follows a 3D Gaussian function, with the standard deviation $\sigma_{\rm HI}=120\,$pc, on top of a constant ground, i.e., 
\begin{equation}\label{eq:HIdensity}
    n_{\rm HI}(x,y,z)=\frac{M_{\rm HI}}{(2\pi)^{3/2}\sigma_{\rm HI}^3m_{\rm H}}\exp\left\{-[x^2+y^2+z^2]/2\sigma_{\rm HI}^2\right\}+n_0
\end{equation}
where we set a Cartesian coordinate system with the origin at the centre of the CR source, the x-axis following the line-of-sight (LOS) of the observer to the CR source, the y-axis along the line connecting the centre of CR source and the centre of the HI gas, and the z-axis being perpendicular to the page pointing to readers. The center of the HI gas is assumed to be at the same distance of the CR source from us. The parameters $M_{\rm HI}$ and $n_0$ are determined by reproducing two observational facts: (1) the total mass of HI gas in the bubble inferred, i.e., $m_{\rm H}\int n_{\rm HI}dxdydz=5.9\times 10^6M_\odot$ based on observation where $m_{\rm H}$ is the mass of proton; (2) the 1D radial distribution of the HI column density as shown in Fig.~\ref{fig:HIgas}. It results in $M_{\rm HI}=4.4\times10^{6}\,M_\odot$ and $n_0\simeq 1.3 \rm \,cm^{-3}$.

The distribution of molecular clouds is more discrete than the atomic gas. The hot spots seen by LHAASO are associated with two big clumps of MCs about $2^\circ$ away from the center of the Cygnus Bubble. In the model we simply consider a spherical cloud with the total mass of $1.2\times 10^6M_\odot$ for MCs according to the mass inferred from the CO emission within the green circle (with a radius of $3^\circ$) in the top-left panel of Fig~\ref{CO_HI_MAP_region}. Its density distribution is also assumed to follow a 3D Gaussian function, with $\sigma_{\rm MC}=25\,$pc since the angular size of the clump
is $\sim 1^\circ$. The projected distances of the CO clumps are about 50\,pc (or an angular separation about $2^\circ$ from the center of the bubble), but this is not necessarily the physical distance of the cloud from the central CR source. In the model, the true distance of the cloud to the CR source is adjusted as a model parameter to reproduce the flux extracted by the CO template, while keeping their projected distances from the CR source to be $y_{\rm MC}=-50\,$pc. The $gamma$-ray emission associated with the CO template outside $3^\circ$ appears rather diffuse without significant clustering. We therefore assume a homogeneous distribution of molecular gas with hydrogen density $n_1$ in addition to the compact cloud. Thus the density distribution is
\begin{equation*}\label{eq:MCdensity}
        n_{\rm MC}(x,y,z)=\frac{M_{\rm MC}}{(2\pi)^{3/2}\sigma_{\rm MC}^3m_{\rm H}}\exp\left\{-[x-x_{\rm MC})^2+(y-y_{\rm MC})^2+z^2]/2\sigma_{\rm MC}^2\right\}+n_1
\end{equation*}
To reproduce the observation that the mass of MC within $3^\circ$ from the source center is $1.2\times 10^6M_\odot$ and the total mass of MC within $10^\circ$ from the source center is $3.2\times 10^6M_\odot$, we set $M_{\rm MC}\approx 1.2\times 10^{6}M_\odot$ and $n_1\approx 1\, \rm cm^{-3}$. As mentioned earlier, the $gamma$-ray flux associated with the inner $3^\circ$ of the CO template is about 40\% of the flux extracted by the entire CO template. To reproduce this result in the model, we found the physical distance of the compact MC cloud need be about 100\,pc from the source center, leading to $x_{\rm MC}=86.6\,$pc. 

Given the proton distribution Eq.~(\ref{eq:crdis}) and the gas distribution Eqs.~(\ref{eq:HIdensity}) and (\ref{eq:MCdensity}), we can obtain the $gamma$-ray emissivity at each spatial point and then integrate over the LOS towards different direction to obtain a 2D intensity map at different energy. The pionic $gamma$-ray flux (and the accompanying neutrino flux) is calculated following the semi-analytical method developed by Kelner et al.\cite{Kelner:2006}. 
We can then further integrate over the solid angle to obtain the total flux and average over the azimuthal angle to get the 1D surface brightness profile. 

In Fig.~\ref{fig:model}, we have compared the model prediction with the broadband $gamma$-ray spectrum measured from the entire bubble, as well as those of different components. The latter includes the spectrum extracted by the compact Gaussian template (the core region), by the extended Gaussian template, and by the CO template, respectively. The core has an extension of $0.38^\circ$ as depicted by the standard deviation of the 2D Gaussian function. We therefore integrate over the emission arising from interactions between protons and the atomic gas within a radius of $30\,$pc, corresponding to the $3\sigma$ extension of the compact Gaussian template (at a nominal distance of 1.46\,kpc), which is also the assumed injection sphere of protons in the model. Then we compare the integrated flux to the flux measured from the core. Similarly, the extended Gaussian component of LHAASO has an extension of roughly $2^\circ$, comparable to that of the Cygnus Cocoon measured by Fermi-LAT. We integrate over the emission within a radius of $150$\,pc from the center, which corresponds to the $3\sigma$ extension of the extended Gaussian component. The obtained flux is consistent with the spectra of Fermi-LAT Cocoon and the Gaussian component. The flux produced by protons and the molecular clumps are compared with that of the hot spots coincident with the CO clumps. As for the total flux from the entire bubble, it consists of the emission from the aforementioned three components and also that produced by interactions between CR sea and the gas in the line of sight of observers covering the entire $6^\circ$ bubble region. We evaluate this background/foreground component in our model assuming the same CR spectra measured at Earth. The corresponding 1D intensity profiles of three energy bands ($2-20$\,TeV, $25-100$\,TeV, $>100$\,TeV) are also compared with observations in the same figure. The SED and the intensity profiles are simultaneously explained. 

The diffusion coefficient $D(E_p)$ is a key model parameter, since it determines the spectrum and density of diffusing particles. Assuming a typical form of power-law for the diffusion coefficient for the entire bubble, i.e, $D(E)=D_0(E_p/E_0)^\delta$, a small diffusion coefficient is needed in order to confine all the injected $\lesssim$TeV protons inside the observed bubble (i.e., $t_{\rm age}<R_b^2/4D(E)$), so as to reproduce the relatively flat spectrum of the Cygnus Cocoon at the GeV band measured by Fermi-LAT (i.e., with a photon index around $2.2$). At higher energies, protons diffuse sufficiently fast and escape to the distance larger than the apparent bubble size, leading to the softening of the spectrum at a few TeV. With this condition, we generally require $D(1\,{\rm TeV})<R_b^2/4t_{\rm age}=9\times 10^{26}(R_b/160{\rm pc})^2(t_{\rm age}/2{\rm Myr})^{-1}\,\rm cm^2/s$. We fit the data with $D_0\simeq 3\times 10^{26}\,\rm cm^2/s$ with $E_0=1\,$TeV and $\delta=0.7$. The diffusion coefficient is about two orders of magnitude smaller than that of the standard interstellar diffusion coefficient which may indicate the operation of the streaming instability, where injected protons are regulated by the self-excited Alfv{\'e}n waves and stream at the speed of Alfv{\'e}n velocity in the plasma. Alternatively, the small diffusion coefficient may reflect the strong turbulence driven by the collective stellar wind from Cygnus OB2. 

\begin{figure}
    \centering
    \includegraphics[width=0.8\textwidth]{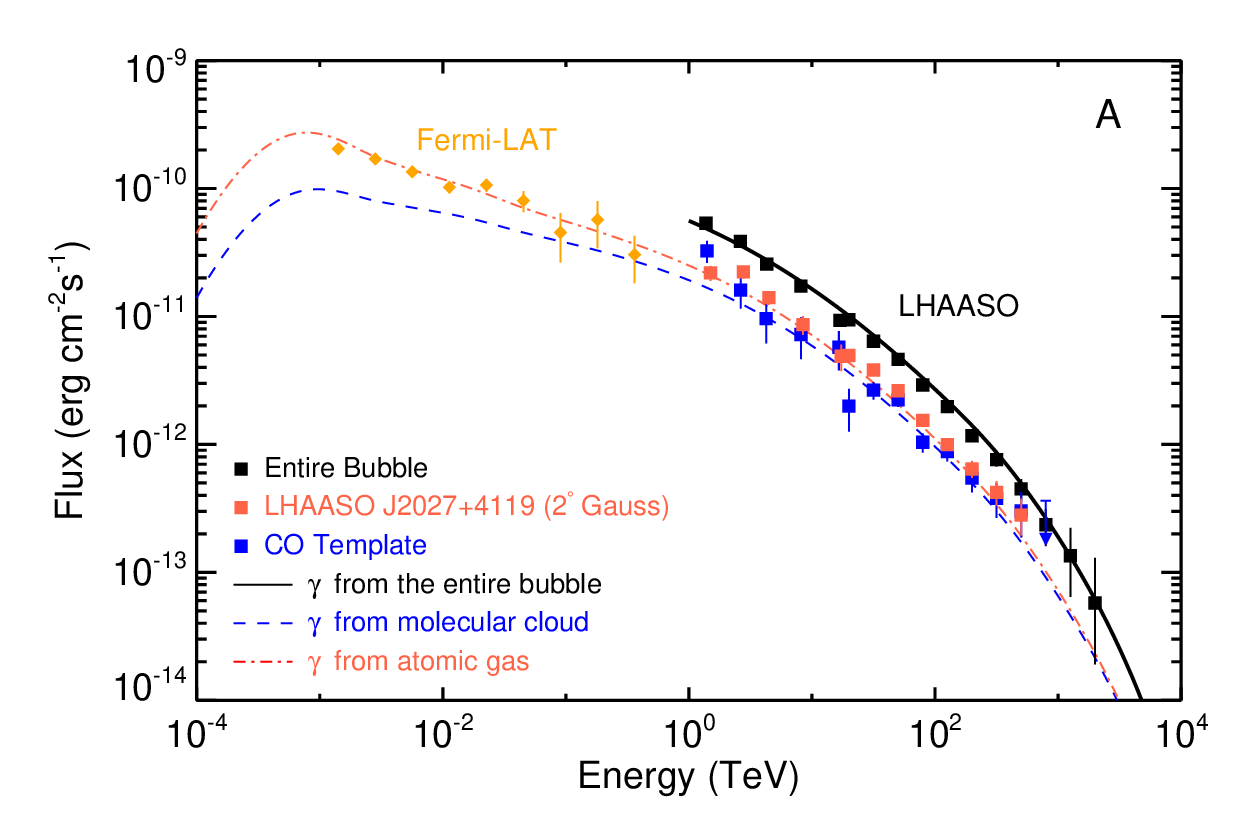}
    \includegraphics[width=0.74\textwidth]{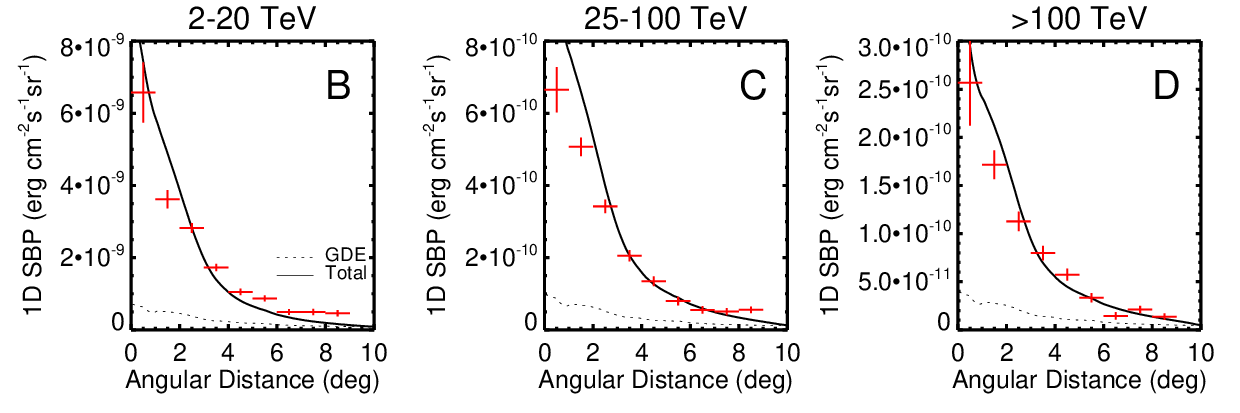}
    \caption{Same as Fig.~\ref{fig:model} but for $D(E)\propto E^{1/3}$. In this case, a broken PL function is employed for the injection proton spectrum, i.e., $N(E_p)=N_0E_p^{-2.25}(1+E_p/30{\rm TeV})^{-0.5}\exp(-E_p/10{\rm PeV})$. The employed diffusion coefficient is $D_0=5\times10^{26}\,\rm cm^2/s$. Other parameters are the same with those in Fig~\ref{fig:model}.}
    \label{fig:model_d0.3}
\end{figure}

The employed diffusion coefficient in Fig.~\ref{fig:model} is not exclusive. We may fit the data with other combinations of $D_0$ and $\delta$. In particular,  a larger available parameter space could be achieved if we relax the presumed power-law form of the diffusion coefficient or the injection spectrum. As is shown in Fig.~\ref{fig:model_d0.3}, for example, we may introduce a spectral break at the injection spectrum around $\sim 10\,$TeV to account for the break in the observed $gamma$-ray spectrum at $\sim$TeV for LHAASO~J2027+4119, and fit the data with $D_0=5\times10^{26}\rm cm^2/s$ and $\delta=1/3$, where the latter is consistent with the Kolmogorov-type turbulence. The required proton luminosity is still $10^{37}\rm erg/s$ in this case.

\begin{figure*}
    \centering
    \includegraphics[width=0.32\textwidth]{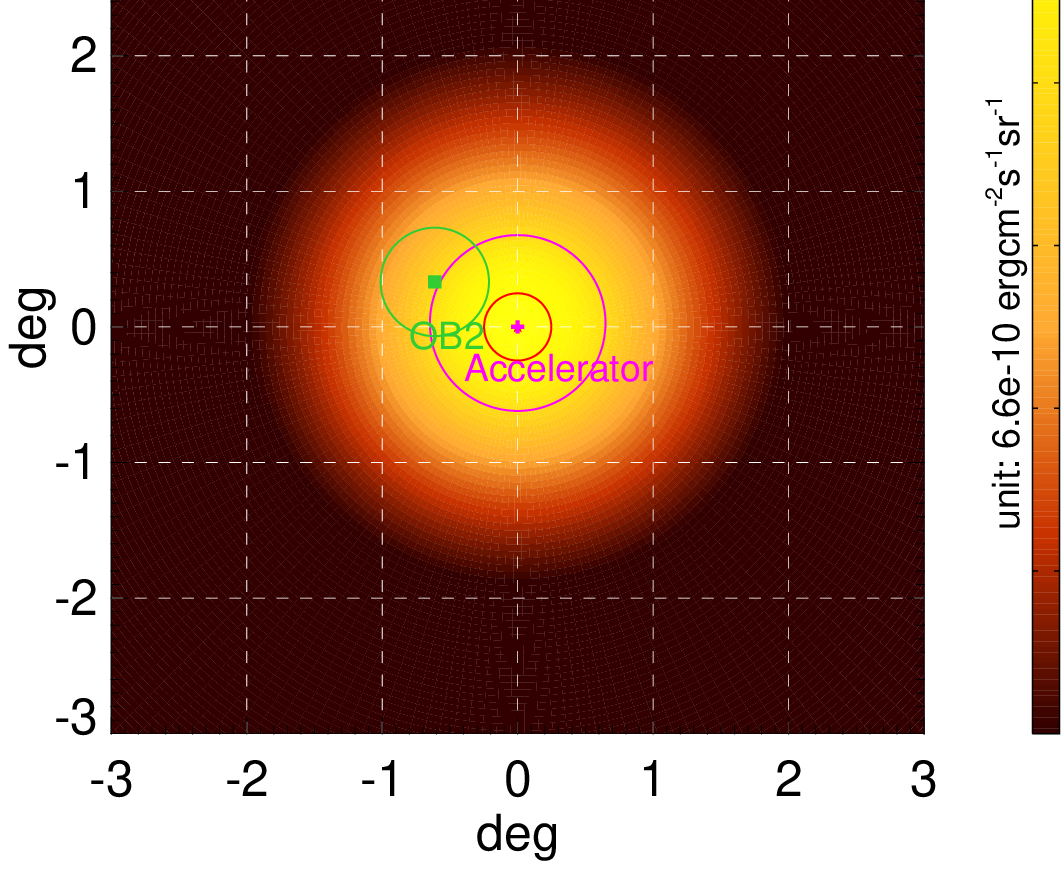}
    \includegraphics[width=0.32\textwidth]{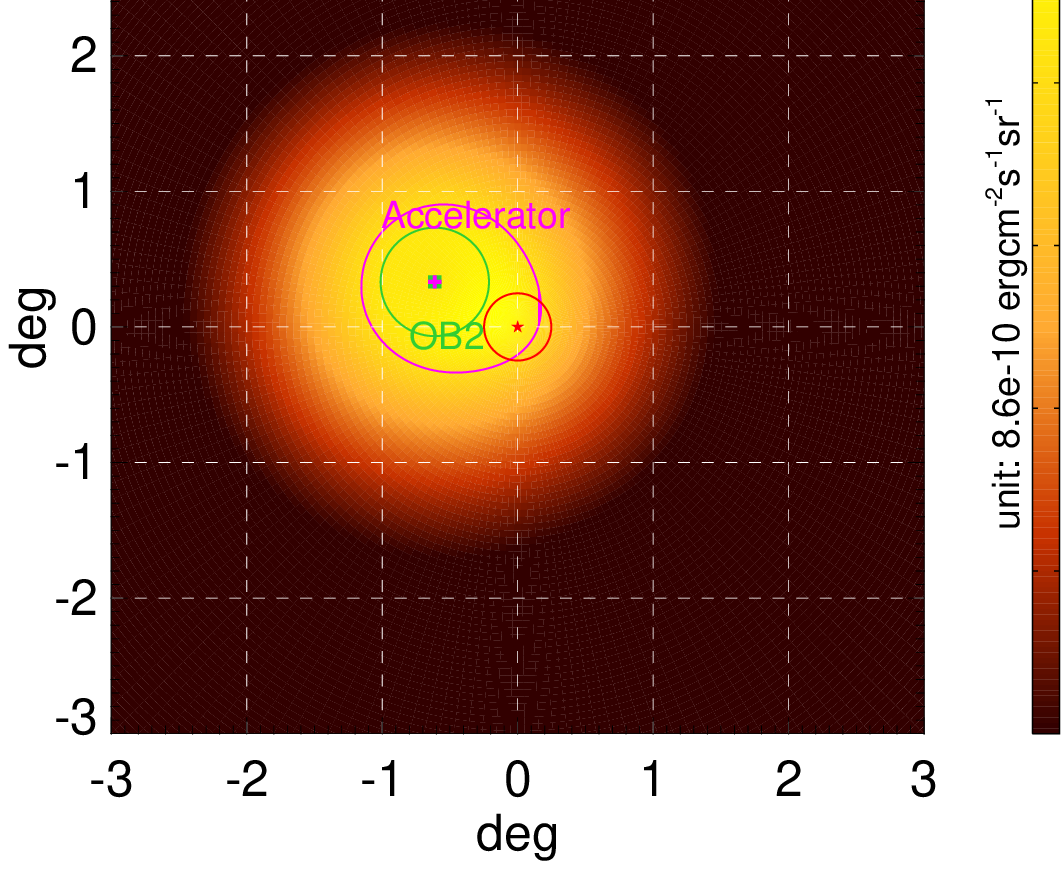}
    \includegraphics[width=0.32\textwidth]{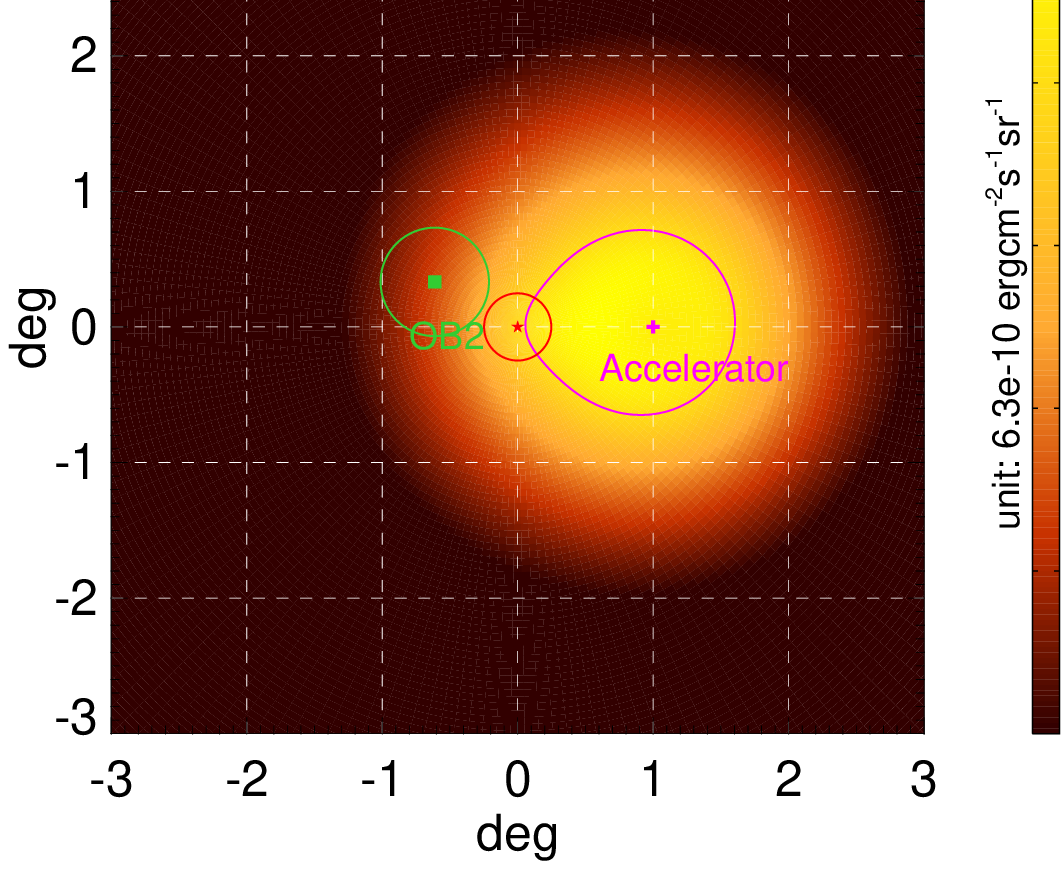}
    \includegraphics[width=0.32\textwidth]{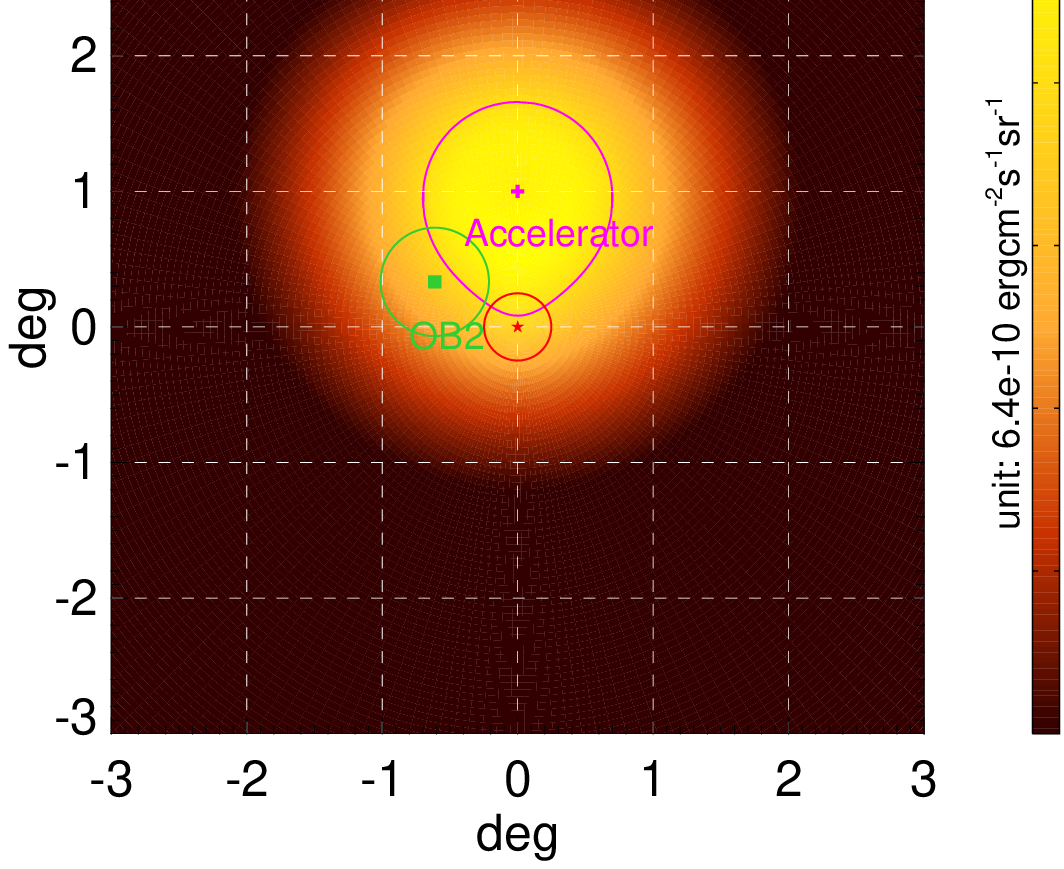}
    \includegraphics[width=0.32\textwidth]{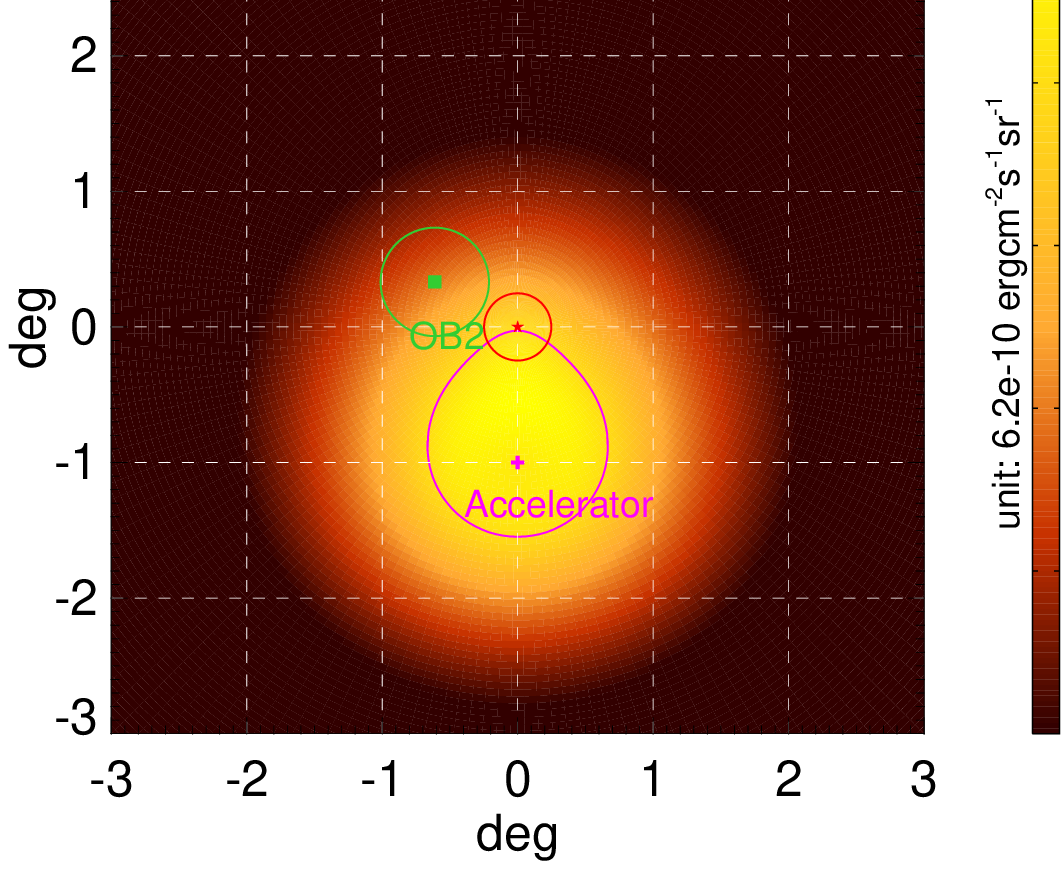}
    \includegraphics[width=0.32\textwidth]{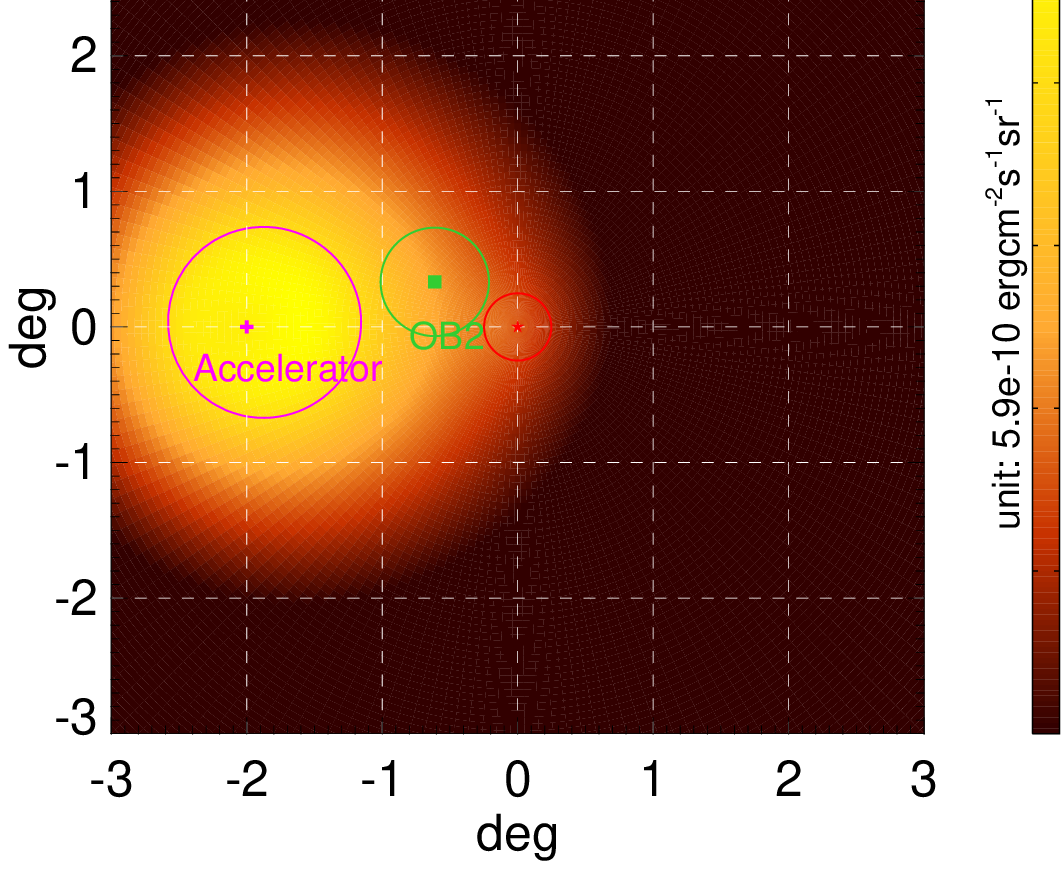}
    \includegraphics[width=0.32\textwidth]{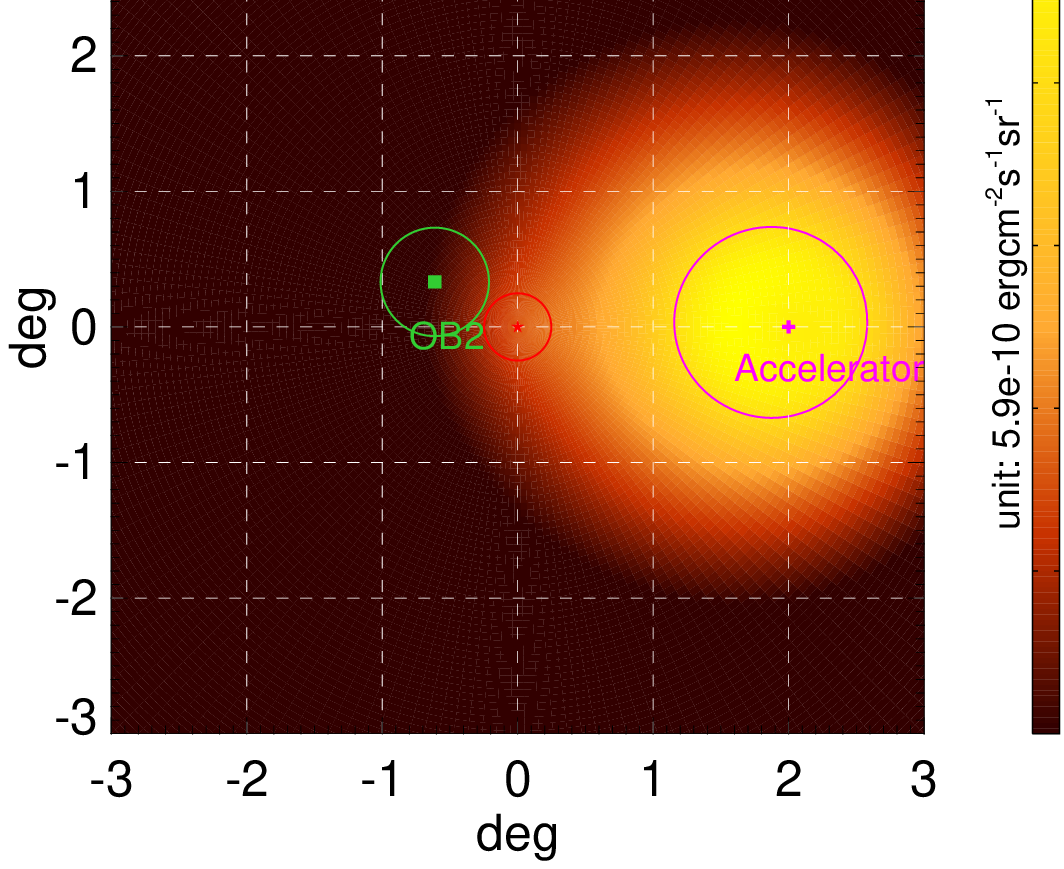}
    \includegraphics[width=0.32\textwidth]{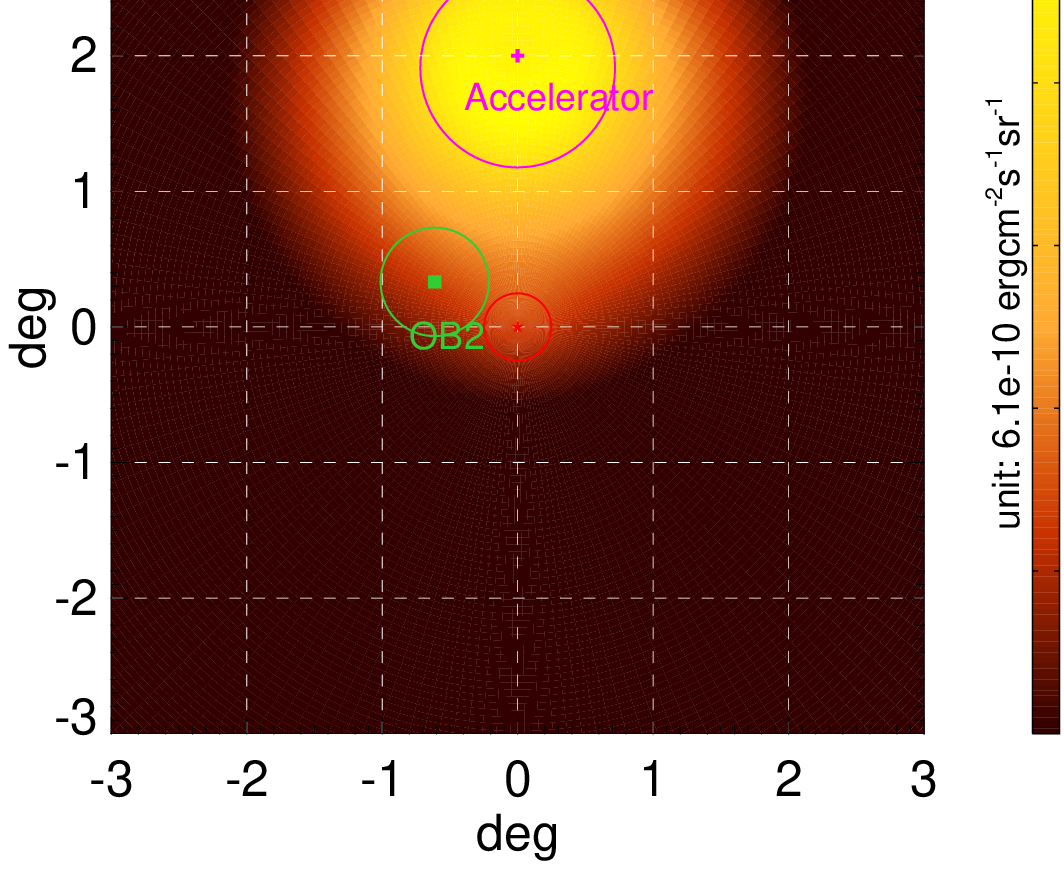}
    \includegraphics[width=0.32\textwidth]{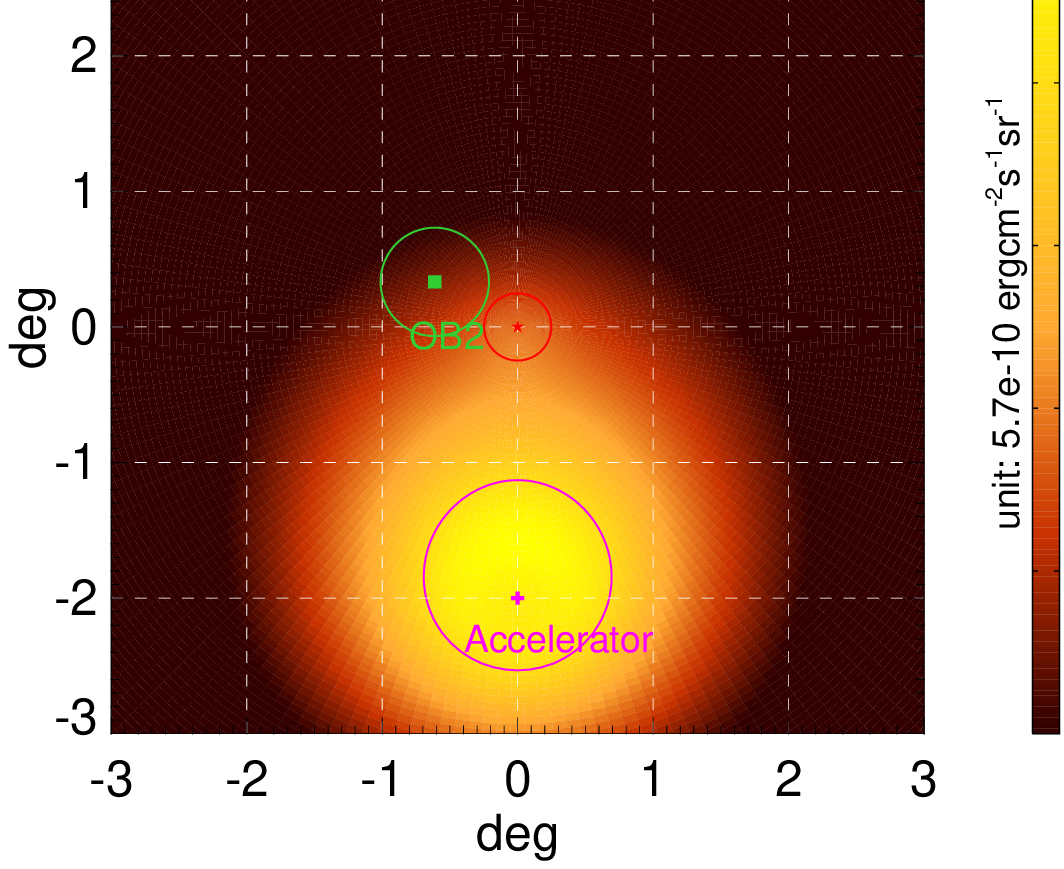}
    \caption{Simulated 2D intensity map from interactions between injected protons and HI gas. The position of the particle accelerator is located at $(x,y,z)=(0,0,0)$, $(0, -8, -15)$, $(0, 0, -25)$, $(0, 25, 0)$, $(0, -25, 0)$, $(0, 0, 50)$, $(0, 0, -50)$, $(0, 50, 0)$, and $(0, -50, 0)$, for the nine panels, respectively, from left to right and top to bottom. The values of the coordinates are in unit of parsec. The red circle represents the error circle of the KM2A center. The magenta circle outlines the brightest part of the emission where the intensity is higher than 90\% of the maximum intensity in the map, and we recognize it as the centroid of the source. The green box and the circle marks the center of Cygus OB2 and the rough size of the starburst region. }
    \label{fig:model_2Dimage}
 \end{figure*}

A key issue is to pinpoint the particle accelerator responsible for the Cygnus Bubble. In our previous calculation, we fix the center of the accelerator to the position spatially coincident with the center of the Gaussian component measured by KM2A in the plane of the sky, at a distance of 1.46\,kpc from the Earth. This corresponds to a coordinate of $(x,y,z)=(0,0,0)$ in the Cartesian coordinate system defined in Fig.~\ref{fig:sketch}. If we shift the position of the particle accelerator a few degrees away, the resultant overall SED and the 1D SBP for the $6^\circ$ bubble will not be changed significantly, but it could lead to a remarkable offset between the predicted centroid of the source and the Gaussian center. In Fig.~\ref{fig:model_2Dimage}, we show the projected 2D intensity map at the energy above 25\,TeV from our model with different position of the particle accelerator. Note that the contribution from molecular clouds are excluded here because the $2^\circ$ Gaussian component measured by LHAASO originate from the hadronic interactions between protons interacting with HI gas in the bubble. In the figure, the red circle represents the error circle of the KM2A center. The magenta circle outlines the brightest part of the emission where the intensity is higher than 90\% of the maximum intensity in the map, and we recognize it as the centroid of the source. The green box and the circle marks the center of Cygnus OB2 and the rough size of the starburst region. When the red circle and the magenta circle overlaps with each other, we consider the model reproduce the position of the observed KM2A center. Based on this criteria, we see from the figure that we may constrain the accelerator within about $1^\circ$ from the current KM2A center, implying the accelerator is located at the core region.  

\section{Corresponding Neutrino Flux and Detection}
If the $gamma$-ray emission of the bubble indeed come from the hadronic interaction of protons, neutrinos will be produced simultaneously in the process via the decay of produced pions. Equal amount of $\pi^0$, $\pi^+$, $\pi^-$ are produced in the $pp$ collisions. Considering the flavor ratio of 1:1:1 after oscillation of neutrinos, same amount of $gamma$-ray photons and (anti-)muon neutrinos will be generated from $pp$ collisions. 
Based on the semi-analytical method provided by Kelner et al.\cite{Kelner:2006}, we can obtain the (anti-)muon neutrino flux from the Cygnus Bubble $dN_\nu/dE_{\nu_\mu}$. The number of track-like events in IceCube expected from the bubble can then be estimated by $N_{\nu_\mu}=\int A_{\rm eff}(E_\nu)(dN_\nu/dE_{\nu_\mu})dE_\nu$, where
$A_{\rm eff}$ is the effective area of IceCube for (anti-)muon neutrino for declination of the Cygnus region \cite{IceCube2020}.
For the parameter set employed in Fig.~\ref{fig:model}, we can expect detection of 29/7.7/0.36 (anti-)muon neutrino events above 1/10/100\,TeV, by IceCube for ten-year operation, where about 39\%/33\%/30\% of them comes from the inner $3^\circ$ region.

\begin{figure}
    \centering
    \includegraphics[width=0.45\textwidth]{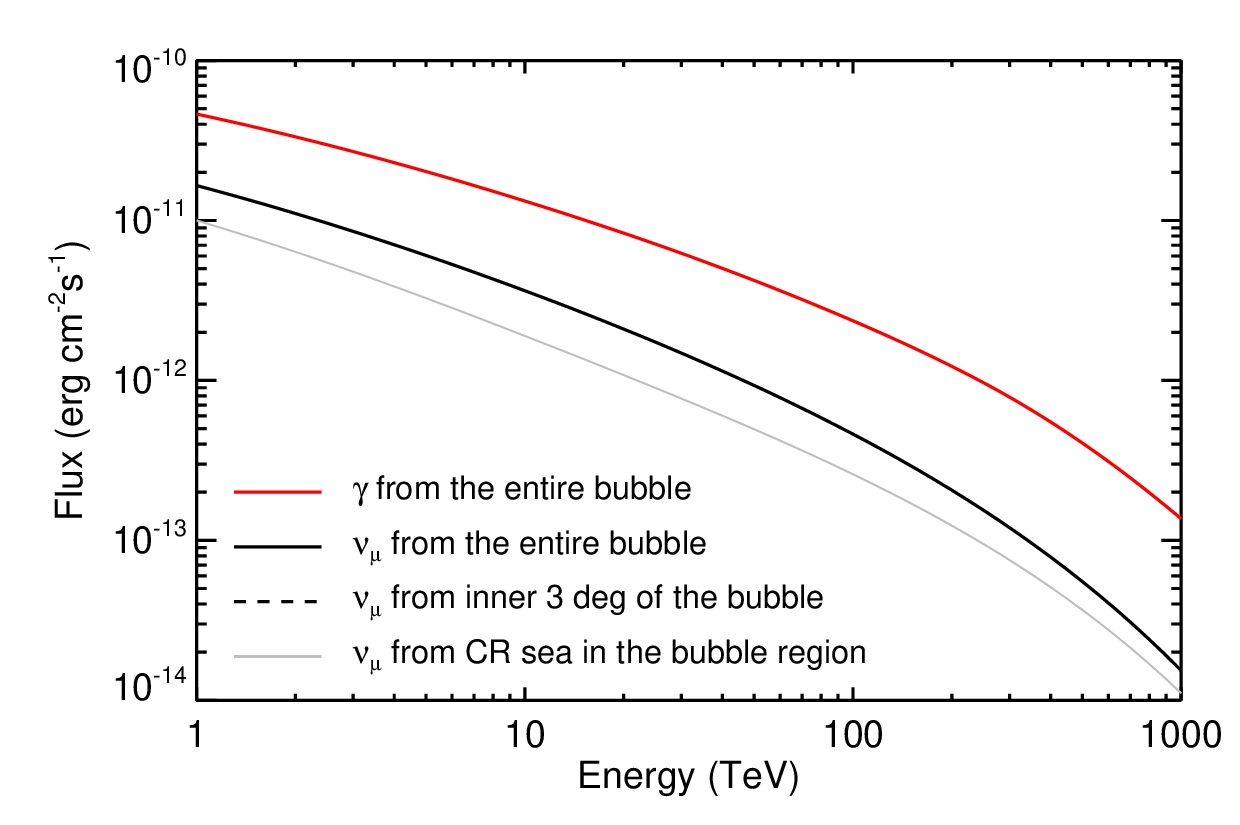}
    \caption{Flux of gamma rays and (anti-)muon neutrinos from the Cygnus region. The black solid curve and dashed curve represent the (anti-)muon neutrino fluxes from the entire bubble and inner 3 degree of the bubble respectively (produced by source accelerated protons only), based on the model shown in Fig.~\ref{fig:model}. The red solid curve represents the $gamma$-ray from the entire bubble. The grey solid curve shows the expected (anti-)neutrino flux from the CR sea in the entire bubble region based on the locally measured CR flux and the gas column density in the bubble region inferred from Planck's observation on the dust emission. Model parameters are the same with those in Fig.~\ref{fig:model}.}
    \label{fig:gamma_nu}
\end{figure}

\end{document}